\pdfoutput=1

\documentclass[reprint,aps,prl,twocolumn,floatfix,superscriptaddress]{revtex4-1}
\usepackage{graphicx}
\usepackage{amssymb}
\usepackage{dcolumn}
\usepackage{bm}
\usepackage{amsmath, siunitx}
\usepackage{lineno}
\usepackage[english]{babel}
\usepackage[utf8]{inputenc}
\usepackage{xcolor}

\usepackage{blindtext}
\usepackage{textcomp,gensymb}
\usepackage[hidelinks]{hyperref}
\allowdisplaybreaks
\begin{document}

\title{Hard superconducting gap in germanium}

\author{Alberto Tosato}
\affiliation{QuTech and Kavli Institute of Nanoscience, Delft University of Technology, PO Box 5046, 2600 GA Delft, The Netherlands}
\author{Vukan Levajac}
\affiliation{QuTech and Kavli Institute of Nanoscience, Delft University of Technology, PO Box 5046, 2600 GA Delft, The Netherlands}
\author{Ji-Yin Wang}
\affiliation{QuTech and Kavli Institute of Nanoscience, Delft University of Technology, PO Box 5046, 2600 GA Delft, The Netherlands}
\author{Casper J. Boor }
\affiliation{QuTech and Kavli Institute of Nanoscience, Delft University of Technology, PO Box 5046, 2600 GA Delft, The Netherlands}
\author{Francesco Borsoi}
\affiliation{QuTech and Kavli Institute of Nanoscience, Delft University of Technology, PO Box 5046, 2600 GA Delft, The Netherlands}
\author{Marc Botifoll}
\affiliation{Catalan Institute of Nanoscience and Nanotechnology (ICN2), CSIC and BIST, Campus UAB, Bellaterra, 08193 Barcelona, Catalonia, Spain}
\author{Carla N. Borja}
\affiliation{Catalan Institute of Nanoscience and Nanotechnology (ICN2), CSIC and BIST, Campus UAB, Bellaterra, 08193 Barcelona, Catalonia, Spain}
\author{Sara Martí-Sánchez}
\affiliation{Catalan Institute of Nanoscience and Nanotechnology (ICN2), CSIC and BIST, Campus UAB, Bellaterra, 08193 Barcelona, Catalonia, Spain}
\author{Jordi Arbiol}
\affiliation{Catalan Institute of Nanoscience and Nanotechnology (ICN2), CSIC and BIST, Campus UAB, Bellaterra, 08193 Barcelona, Catalonia, Spain}
\affiliation{ICREA, Passeig Lluís Companys 23, 08010 Barcelona, Catalonia, Spain}
\author{Amir Sammak}
\affiliation{QuTech and Netherlands Organisation for Applied Scientific Research (TNO), Stieltjesweg 1, 2628 CK Delft, The Netherlands}
\author{Menno Veldhorst }
\affiliation{QuTech and Kavli Institute of Nanoscience, Delft University of Technology, PO Box 5046, 2600 GA Delft, The Netherlands}
\author{Giordano Scappucci}
\email{g.scappucci@tudelft.nl}
\affiliation{QuTech and Kavli Institute of Nanoscience, Delft University of Technology, PO Box 5046, 2600 GA Delft, The Netherlands}

\date{\today}

\begin{abstract}
The co-integration of spin, superconducting, and topological systems is emerging as an exciting pathway for scalable and high-fidelity quantum information technology. High-mobility planar germanium is a front-runner semiconductor for building quantum processors with spin-qubits, but progress with hybrid superconductor-semiconductor devices is hindered because obtaining a superconducting gap free of subgap states (hard gap) has proven difficult. Here we solve this challenge by developing a low-disorder, oxide-free interface between high-mobility planar germanium and a germanosilicide parent superconductor. This superconducting contact is formed by the thermally-activated solid phase reaction between a metal (Pt) and the semiconductor heterostructure (Ge/SiGe). Electrical characterization reveals near-unity transparency in Josephson junctions and, importantly, a hard induced superconducting gap in quantum point contacts. Furthermore, we demonstrate phase control of a Josephson junction and study transport in a gated two-dimensional superconductor–semiconductor array towards scalable architectures. These results expand the quantum technology toolbox in germanium and provide new avenues for exploring monolithic superconductor-semiconductor quantum circuits towards scalable quantum information processing. 
\end{abstract}

\maketitle

The intimate coupling between superconductors and semiconductors in hybrid devices is at the heart of exciting pursuits, including topological qubits with Majorana zero modes~\cite{Lutchyn2018MajoranaHeterostructures,flensberg_engineered_2021}, superconducting (Andreev) spin qubits~\cite{hays_coherent_2021}, and gate-tunable superconducting qubits~\cite{Casparis2018SuperconductingGas}. Combining hybrid devices with high-fidelity semiconductor spin qubits in a single material platform may resolve key challenges for scalable quantum information processing. In particular, quantum information transfer between spin and topological qubits~\cite{choi_spin-dependent_2000,Leijnse2011QuantumSystems,Leijnse2012HybridGates,hoffman_universal_2016} may enable a universal gate set for topological quantum computation and, conversely, superconductors may be used to coherently couple spin qubits at a distance via crossed Andreev reflection~\cite{choi_spin-dependent_2000,Leijnse2013CouplingSuperconductors} or topologically protected links~\cite{Kitaev2001UnpairedWires}.

The use of epitaxial superconducting Al to induce a hard superconducting gap in III-V semiconductors~\cite{Chang2015HardNanowires,Kjaergaard2016QuantizedHeterostructure} stimulated great progress with hybrid devices, leading to experimental reports of topological superconductivity in planar Josephson junctions~\cite{fornieri_evidence_2019} and in electrostatically defined quasi-1D wires~\cite{aghaee_inas-hybrid_2022}, the demonstration of Andreev spin qubits~\cite{hays_coherent_2021}, and the realization of a minimal Kitaev chain in
coupled quantum dots~\cite{dvir_realization_2022}. However, spin qubits in III-V semiconductors suffer from the hyperfine interactions with the nuclear spin bath~\cite{Cywinski2009ElectronBath} that severely deteriorate their quantum coherence~\cite{stano_review_2022} and challenges their integration with hybrid devices. 

On the other hand, spin qubits with quantum dots in Ge~\cite{Scappucci2020TheRoute,Watzinger2018AQubit,Hendrickx2020FastGermanium,Jirovec2021AGe} can achieve long quantum coherence due to the suppressed hyperfine interaction~\cite{Prechtel2016DecouplingSpins} and the possibility of isotopic purification into a nuclear spin-free material~\cite{Itoh2014IsotopeApplications}. Thanks to the light effective mass~\cite{Lodari2019LightWells} and high mobility exceeding one million cm$^2$/Vs~\cite{Lodari2022LightlyMillion}, holes in planar Ge/SiGe heterostructures have advanced semiconductors spin qubits to the universal operation on a $2\times2$ qubit array~\cite{Hendrickx2021AProcessor}, and the shared control of a 16 semiconductor quantum dot crossbar array~\cite{borsoi_shared_2022}.
Moreover, the ability of holes to make contacts with low Schottky barrier heights to metals~\cite{Dimoulas2006Fermi-levelGermanium}, including superconductors, makes 
Ge a promising candidate for hybrid devices. 
Initial work used superconducting Al to contact Ge either via thermal diffusion~\cite{Hendrickx2018Gate-controlledGermanium,Hendrickx2019BallisticGermanium,Ridderbos2019HardNanowires} or by deposition on the sidewalls of etched mesas~\cite{Vigneau2019GermaniumInterferometers, Aggarwal2021EnhancementGas}. However, the key demonstration of a superconducting gap in Ge free of subgap quasiparticle states is lacking, challenged by the difficulty of contacting uniformly a buried quantum well with a superconductor, whilst maintaining the low disorder at the superconductor-semiconductor interface and in the semiconductor channel.

Here we solve these challenges and demonstrate a hard superconducting gap in Ge. We contact the quantum well with a superconducting germanosilicide (PtSiGe), similar to the silicidation process used by the microelectronics industry for low resistance contacts~\cite{Kittl2008SilicidesApplications}. The superconductor is formed uniformly within the heterostructure and reaches the buried quantum well via a controlled thermally-activated solid phase reaction between the metal (Pt) and the semiconductor stack (Ge/SiGe). This process is simple, robust, and does not require specialised vacuum conditions or etching because the superconductor-semiconductor interface is buried into the pure semiconducting heterostructure and consequently remains pristine. This represents a paradigm shift compared to the subtractive nanofabrication processes commonly used for hybrid devices, since our additive process does not deteriorate the active area of the semiconductor.
As a result, we demonstrate a suite of reproducible Ge hybrid devices with low disorder and excellent superconducting properties. 

\begin{figure*}[!ht]
	\includegraphics[width=175mm]{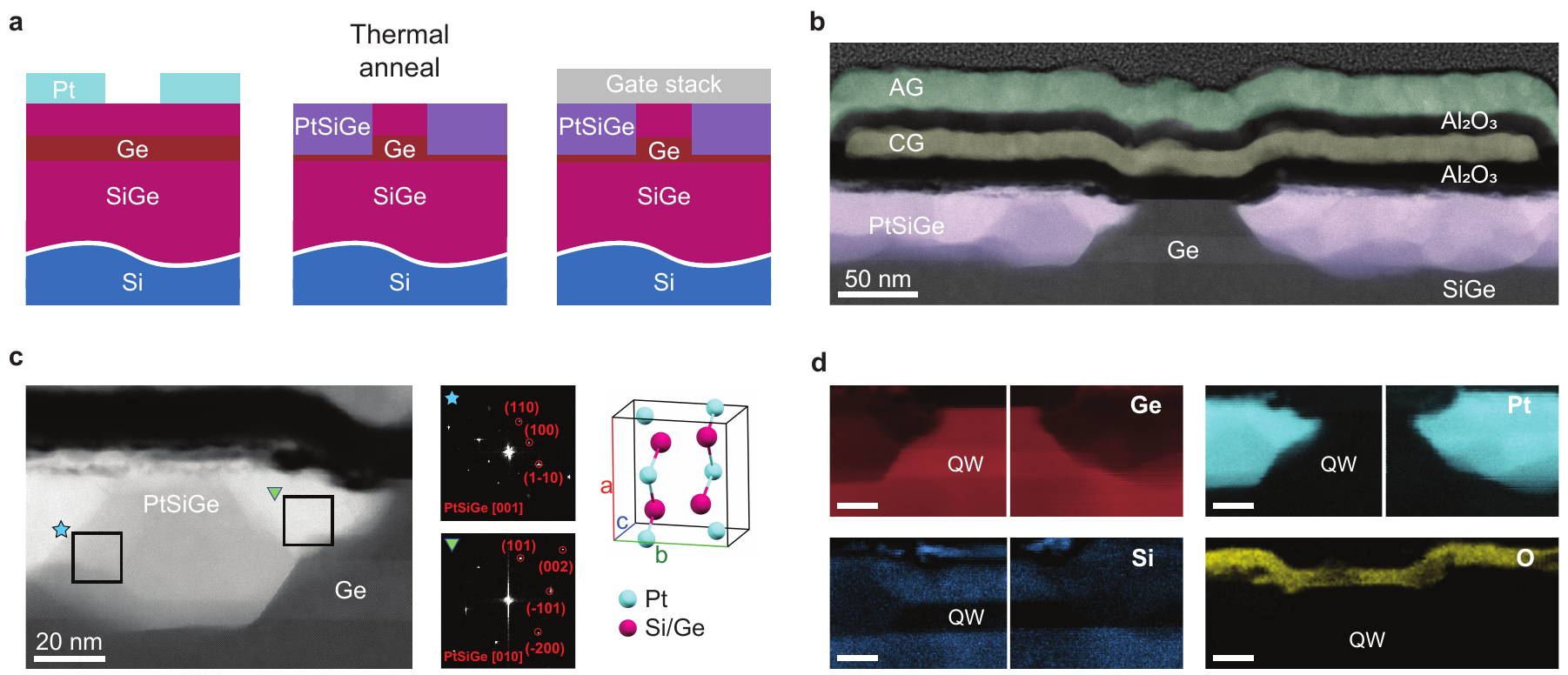}
	\caption{\textbf{Material properties of superconductor-semiconductor Ge devices.} \textbf{a}) Schematics of the fabrication process for a superconductor-normal-superconductor quantum point contact (SNS-QPC). First, platinum is deposited on the heterostructure, then thermal annealing at \SI{400}{^\degree C} drives Pt in the heterostructure to form PtSiGe, finally two gate layers are deposited, insulated by Al$_2$O$_3$. \textbf{b}) False-color high angle annular dark field scanning transmission electron microscopy (HAADF STEM) image of a cross-section of a SNS-QPC. The PtSiGe contacts are violet, the Ti/Pd constriction gate (CG) operated in depletion mode is yellow, the Ti/Pd accumulation gate (AG), used to populate the quantum well, is green. A scanning electron microscopy top view image of this device is shown in Fig.~\ref{fig:SNS}. \textbf{c}) Atomic resolution HAADF STEM image of the Ge/PtSiGe interface along with the indexed fast Fourier transforms (FFTs) of the two regions (black squares) within the PtSiGe contacts and a schematics of the PtSiGe orthorhombic unit cell. The corresponding ternary lattice parameters $T = a_T, b_T, c_T$ that define the dimensions of the unit cell can be calculated, in a first approximation, by Vegard’s law: $T_{\mathrm{PtSi_{1-x} Ge_x}}=x\ B_{\mathrm{PtGe}} + (1-x)\ B_{\mathrm{PtSi}}$ where $B = a_B, b_B, c_B$ are the lattice parameters of the binary compounds PtSi and PtGe, and $x$ is the relative content of Ge with respect to Si. \textbf{d}) Electron energy-loss spectroscopy (EELS) composition maps showing the Pt, Ge, Si and O signals for the central area of the TEM lamella of panel b, the scale-bar indicates \SI{50}{nm}. The PtGeSi stoichiometry is extracted by quantitative EELS analysis and reported in Supplementary Fig.~S4}.
\label{fig:materials}
\end{figure*}

\begin{figure*}[!ht]
	\includegraphics[width=175mm]{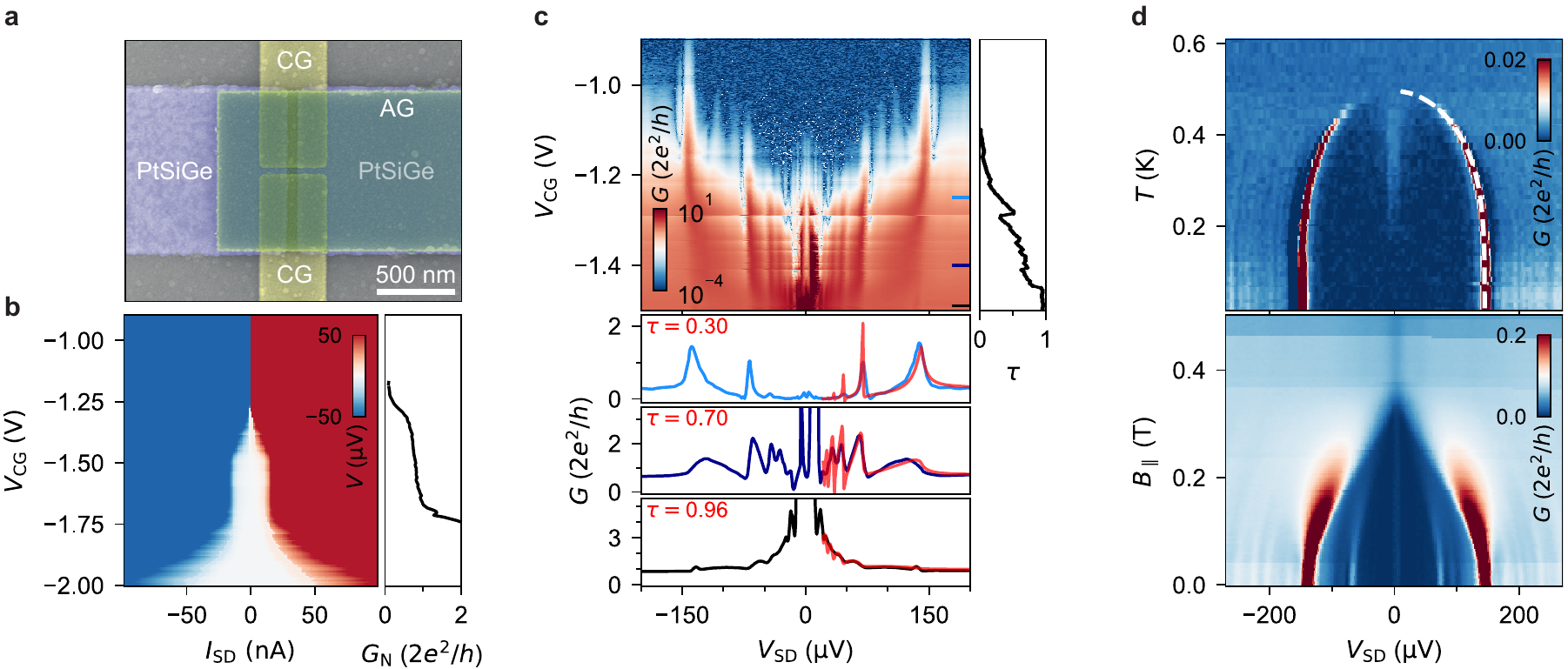}
	\caption{
        \textbf{Highly-transparent Josephson junctions.}
        \textbf{a}) False-color scanning electron microscope image of the SNS device. The PtSiGe contacts are violet, the constriction gates (CG) are yellow and the accumulation gate (AG) is green. The channel length between the two superconducting leads is \SI{70}{nm} and the channel width between the constriction gates is \SI{40}{nm}. The two constriction gates are separate by design but always shorted together during measurements. 
        \textbf{b}) Color map of the voltage drop across the junction $V$ \emph{vs} source-drain current $I_\mathrm{SD}$ and constriction-gate voltage $V_\mathrm{CG}$ along with normal-state conductance ($G_\mathrm{N}$) trace \emph{vs} $V_\mathrm{CG}$. $G_\mathrm{N}$ is calculated as the conductance average where the voltage drop across the device is in the range [500, 650]~mV or -[650,500]~mV, that is much higher than the estimated superconducting gap. \textbf{c}) Color map of $G$ in units of $2e^2/h$ \emph{vs} the source-drain voltage $V_\mathrm{SD}$ and $V_\mathrm{CG}$. Bottom panel shows line-cuts of conductance at $V_\mathrm{CG} = [-1.25, -1.4, -1.49]\ \mathrm{V}$, red lines are the fit with the coherent scattering model from which transparency $\tau$ is extracted. Right inset shows the evolution of the transparency, as extracted from the fitting of conductance curves to the coherent scattering model (Methods), with the constriction gate $V_\mathrm{CG}$. \textbf{d}) Color map of $G$ \emph{vs} $T$ and $V_\mathrm{SD}$ (top panel), and \emph{vs} $B_{\parallel}$ and $V_\mathrm{SD}$ (bottom panel), where $B_{\parallel}$ is the in-plane magnetic field in the direction of transport and $T$ the temperature. The color scale in panel (d) has been saturated to better infer the low conductance limit. The source-drain bias is applied between the PtSiGe contacts, and the voltage drop across the junction is measured with a standard 4-terminal setup. The accumulation voltage for measurements in b,c and d was set to \SI{-4.5}{V}, where the 2DHG is expected to reach saturation density (see Supplementary Fig.~S1) of $\simeq 6\times 10^{11}$~cm$^{-2}$~\cite{Sammak2019ShallowTechnology}. Measurement presented in b, c and in panel d (bottom), are performed at \SI{15}{mK}, corresponding to an electron temperature of $\sim\SI{25}{mK}$.
	}
\label{fig:SNS}
\end{figure*}
\subsection{Material properties}
Our approach to superconductor-semiconductor hybrid devices in Ge is illustrated in Fig.~\ref{fig:materials}a. We use an undoped and compressively-strained Ge quantum well, grown by chemical vapor deposition on a Si(001) wafer~\cite{Sammak2019ShallowTechnology} and separated from the surface by a SiGe barrier (Methods). This heterostructure supports a two-dimensional hole gas (2DHG) with high mobility ($\sim 6\times10^5 \ \mathrm{cm^2/Vs}$), long transport scattering time $\tau$ ($\sim 30\ \mathrm{ps}$), and long mean free path ($\sim\SI{7}{\micro m}$) (Supplementary Fig.~S1) and hosts high-performance spin-qubits~\cite{Hendrickx2020FastGermanium}. Crucial for the reliable search of topological superconductivity~\cite{Ahn2021EstimatingNanowires} and for scaling to large spin-qubit architectures~\cite{Vandersypen2017InterfacingCoherent}, the disorder in our buried Ge quantum wells is characterised by an energy level broadening $\hbar/2\tau$ of
$\sim 0.01$~meV, which is more than one order of magnitude smaller than in the other material systems exhibiting a hard superconducting gap~(Supplementary Table~S1).

As shown by the schematics in Fig.~\ref{fig:materials}a, we obtain PtSiGe contacts to the quantum well by room-temperature evaporation of a Pt supply layer, metal lift-off, and rapid thermal process at \SI{400}{\degree C} (Methods). This low-temperature process preserves the structural integrity of the quantum well grown at \SI{500}{\degree C}, whilst activating the solid phase reaction driving Pt into the heterostructure and Ge and Si into the Pt (Supplementary Fig.~S3). As a result, low-resistivity germanosilicide phases are formed~\cite{Gaudet2006ThinGermanium} and under these process conditions the obtained PtSiGe films are superconducting with a $T_\mathrm{c}\approx\SI{0.5}{K}$ and an in-plane critical field of $B_\mathrm{c\|}\approx 400$~mT {(Supplementary Fig.~S2)}. Finally, we use patterned electrostatic gates, insulated by dielectric films in between, to accumulate charge carriers in the quantum well and to shape the electrostatic confinement potential of the hybrid superconductor-semiconductor devices (Methods). Because we do not perform any etch during the nanofabrication of hybrid devices, the low-disorder landscape that determines the 2DHG high mobility is likely to be preserved when further dimensional confinement is achieved by means of electrostatic gates. This is different for hybrid devices with III-V semiconductors, since etching of the superconductor in the active semiconductor region causes mobility degradation~\cite{Shabani2016Two-dimensionalNetworks}.

The morphological, structural, and chemical properties of the hybrid devices are inferred by aberration corrected (AC) high-angle annular dark-field scanning transmission electron microscopy (HAADF-STEM) and electron energy-loss spectroscopy (EELS). Fig.~\ref{fig:materials}b shows a HAADF-STEM image of a cross-section of a superconductor-normal-superconductor quantum point contact (SNS-QPC) taken off-center to visualise the two gate layers (Fig~\ref{fig:SNS}a shows a top view of the device). We observe a uniform quantum well of high-crystalline quality, with sharp interfaces to the adjacent SiGe and absence of extended defects. As a result of the annealing, Pt diffuses predominantly vertically through the SiGe spacer reaching the quantum well. The sharp lateral interfaces between the two PtSiGe contacts and the QW in between set the length of the channel populated by holes via the top-gates. The PtSiGe film presents poly-crystalline domains with a crystal size up to 50$\times$50 nm and orthorhombic phase (PBNM, space group number 62)~\cite{Alptekin2009PlatinumLayers}. This is inferred from the power spectra or fast Fourier transforms (FFTs) taken from the two PtSiGe domains interfacing with the QW from the left contact, shown in Fig.~\ref{fig:materials}c along with a schematic view of the unit cell of such phase.
The analysis of EELS elemental concentration profiles across the Ge QW$\rightarrow$PtSiGe heterointerface (Supplementary Fig.~S4) reveals that the threefold PtSiGe stoichiometry is Ge-rich, with relative composition in the range between $\mathrm{Pt_{0.1}Si_{0.2}Ge_{0.7}}$ and $\mathrm{Pt_{0.1} Si_{0.05} Ge_{0.85}}$ depending locally on the analysed grain.
The EELS compositional maps in Fig.~\ref{fig:materials}d show the elemental distribution of Ge, Si, Pt, Al, and O, at the key regions of the device. We observe Pt well confined to the two contacts areas, which also appear Ge-rich. Crucially, O is detected only in the Al$_2$O$_3$ dielectric layer below the gates, pointing to a high-purity quantum well and a pristine superconductor-semiconductor interface. 

\subsection{Highly transparent Josephson junction}

\begin{figure*}[!ht]
	\includegraphics[width=175mm]{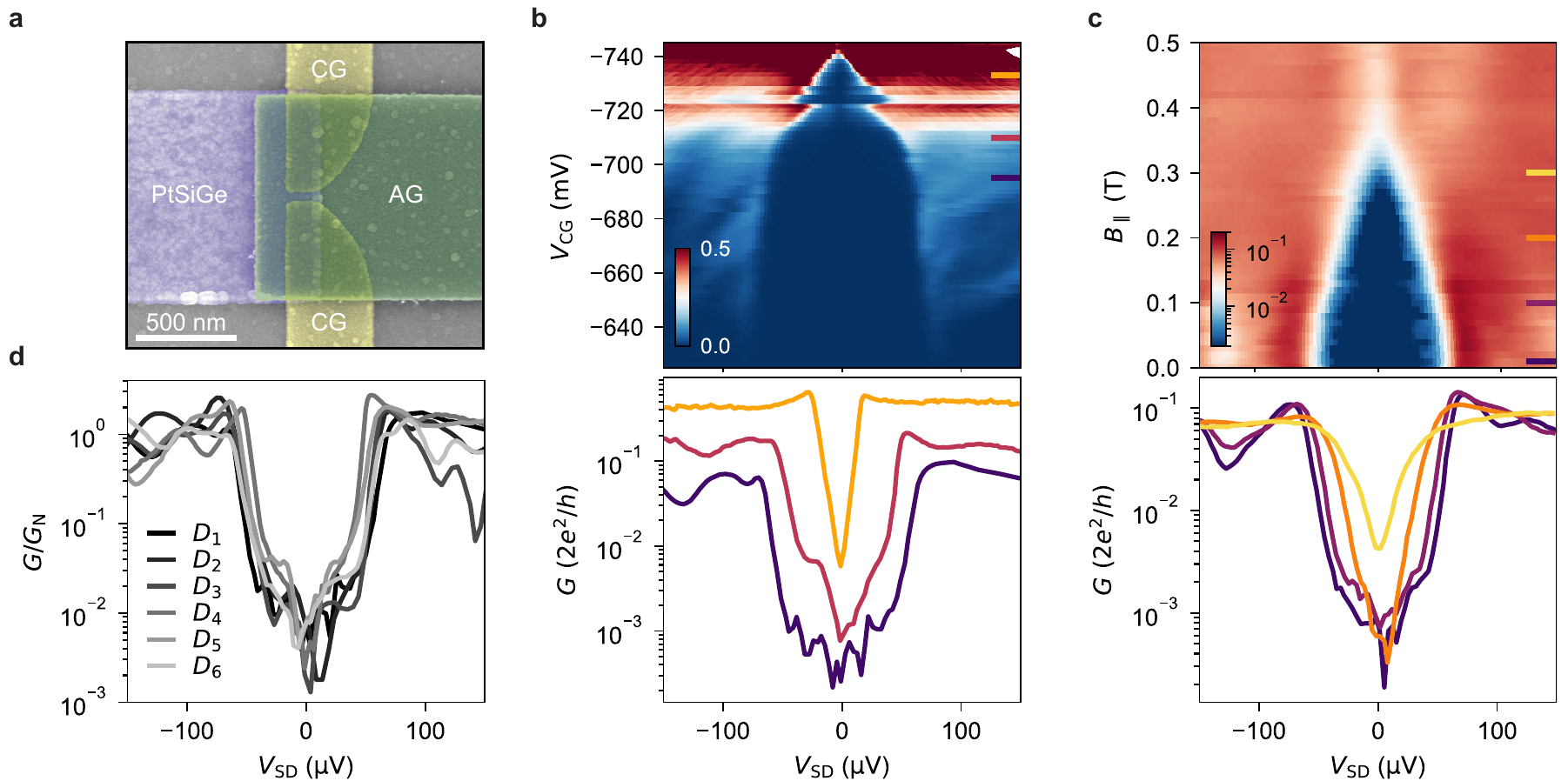}
	\caption{\textbf{Hard induced superconducting gap. a}) False-color SEM image of the superconductor-normal quantum point contact device (SN-QPC). The PtSiGe contact is violet, the constriction gate (CG) are yellow and the accumulation gate (AG) is green. The two constriction gates are separate by design but always shorted together during measurements. \textbf{b}) Color map of conductance $G$ \emph{vs} the source-drain voltage $V_\mathrm{SD}$ and constriction gate $V_\mathrm{CG}$, along with line cuts in log-scale of $G$ at the constriction gate voltages $V_\mathrm{CG}=[-733, -710, -695]$~mV marked by the colored segment in the color-plot. 
	\textbf{c}) Color map of $G$ in units of $2e^2/h$ \emph{vs} the in-plane magnetic field $B_{\parallel}$ perpendicular to the transport direction and constriction gate $V_\mathrm{CG}$, along with line cuts in log-scale of $G$ at the field strength $B_{\parallel}=[0.01, 0.1, 0.2, 0.3]$~T marked by the colored segment in the color-plot.
	\textbf{d}) Conductance traces normalised to the above-gap conductance ($G/G_\mathrm{N}$) \emph{vs} $V_\mathrm{SD}$ in tunneling regime for 6 different SN-QPC devices $D_1$-$D_6$ processed in the same fabrication run, device $D_1$ is the one reported in Fig\ref{fig:SN} a-c, in the remaining devices the constriction gates separation varies (specifications of these devices are provided in Supplementary Fig.~S8).}
\label{fig:SN}
\end{figure*}

We perform low-frequency four-terminal current and voltage bias measurements (Methods) on the SNS-QPC device shown in Fig.~\ref{fig:SNS}a to infer the properties of the superconductor-semiconductor interface. Accumulation (AC, in green) and constriction (CG, in yellow) gates control transport within the 70~nm long channel between the two PtSiGe leads. We apply a large negative voltage to the accumulation gate to populate the quantum well with holes, and we then control the effective width of the channel by applying a more positive voltage to the constriction gates, thus depleting the underlying quantum well. 

The current bias measurements (Fig.~\ref{fig:SNS}b) reveal a tunable supercurrent with a plateau when the constriction gate voltage $V_\mathrm{CG}$ is in the range $\approx [-1.75,-1.50]$~V. This is the same range where we observe the first conductance plateau in the normal-state conductance $G_\mathrm{N}$ (Fig.~\ref{fig:SNS}b, right inset), indicating that the switching current ($I_\mathrm{sw}$) plateau observed in the color plot stems form the supercurrent discretization due to the discrete number of modes in the QPC~\cite{Hendrickx2019BallisticGermanium, Irie2014JosephsonContacts}. Supercurrent discretization up to the third conductance plateau is shown in Supplementary Fig.~S5 (data are for a different SNS-QPC device with identical design to the one presented here). We use the switching current as a lower bound for the critical current and we estimate an $I_{\mathrm{sw}}R_\mathrm{N}$ product of \SI{51}{\micro V}, showing an improvement as compared to previous results obtained with pure Al contacts in Ge QWs~\cite{Hendrickx2019BallisticGermanium, Vigneau2019GermaniumInterferometers, Aggarwal2021EnhancementGas}, despite the Al $T_\mathrm{c}$ is higher than the PtSiGe $T_\mathrm{c}$. The measured $I_\mathrm{sw}R_\mathrm{N}$ product is $\sim 0.5$ the theoretical $I_\mathrm{c}R_\mathrm{N}$ product calculated for a ballistic short junction using the Ambegaokar–Baratoff formula
$\pi\Delta^*/2e = \SI{110}{\micro V}$ with $I_\mathrm{c}$ being the critical current, $\Delta^*$ the induced superconducting gap and $e$ the electron charge~\cite{Ambegaokar1963TunnelingSuperconductors}. This discrepancy has been observed in previous works~\cite{Heedt2021Shadow-wallDevices, Hendrickx2019BallisticGermanium} and is consistent with a premature switching due to thermal activation~\cite{Tinkham1996IntroductionSuperconductivity}.

By operating the device in voltage-bias configuration and stepping the constriction gates, we observe in the conductance color plot the typical signature of multiple Andreev reflections (MARs) (Fig.~\ref{fig:SNS}c). When the applied voltage bias corresponds to an integer fraction of $2\Delta^*$, with $\Delta^*$ being the induced superconducting gap, we observe differential conductance $dI/dV$ peaks (dips) in the tunneling (open) regime~\cite{Octavio1983SubharmonicConstrictions, Kjaergaard2017TransparentJunction}. We measure MARs up to the 5th order, suggesting that the coherence length $\xi_\mathrm{N}$ in the Ge QW is a few times larger than the junction length $L$, and setting a lower bound to the phase coherence length in the QW $l_\psi > 5L=\SI{350}{nm}$. These observations are consistent with the findings of ref.~\cite{Aggarwal2021EnhancementGas} where a similar Ge/SiGe heterostructure is used. Fitting the differential conductance with the coherent scattering model described in ref.~\cite{Averin1995AcChannel} (and used in refs.~\cite{Kjaergaard2017TransparentJunction,Heedt2021Shadow-wallDevices, Borsoi2021SingleShotDevices}) reveals single channel transport with gate tunable transparency up to 96\%. Such a high transparency confirms the high quality interface between the PtSiGe and the Ge QW. From the MARs fit we estimate an induced superconducting gap $\Delta^*=70.6 \pm \SI{0.9}{\micro eV}$, which is about half compared to the $\Delta^*=\SI{129}{\micro eV}$~\cite{aghaee_inas-hybrid_2022} and \SI{150}{\micro eV}~\cite{Fornieri2019EvidenceJunctions} for recent InAs-Al devices reporting topological superconductivity.

Further, we characterise the evolution of the induced superconducting gap with temperature and magnetic field. After setting the device in tunneling regime, where sharp coherence peaks are expected at $e|V_\mathrm{SD}|=2\Delta^*$ (Fig.~\ref{fig:SNS}d), we observe the induced superconducting gap closing with increasing temperature and magnetic field. By fitting the temperature dependence of the coherence peaks with the empirical formula from ref.~\cite{Nilsson2012SupercurrentJunction} we obtain a critical temperature of \SI{0.5}{K}. The peak close to zero bias emerging at $T > \SI{0.2}{K}$ can be explained in terms of thermally-activated quasiparticle current~\cite{Borsoi2021SingleShotDevices}. The in-plane magnetic field in the transport direction quenches the superconductivity at $B_\mathrm{c\|}=\SI{0.37}{T}$. The same critical field is found for the in-plane direction perpendicular to the transport direction while for the out of plane direction $B_\mathrm{c\perp}=\SI{0.1}{T}$ (Supplementary Fig.~S6). This in-plane \emph{vs} out-of-plane anisotropy is expected given the thin-film nature of the PtSiGe superconductor~\cite{Tinkham1996IntroductionSuperconductivity}.

\subsection{Hard induced superconducting gap}

To gain insights into the quality of the Ge/PtSiGe junction we characterise transport through the superconductor-normal quantum point contact (SN-QPC) device shown in Fig.~\ref{fig:SN}a. Importantly, the methodology based on spectroscopy of SN devices alleviates the ambiguity of measuring the amount of quasiparticle states inside the gap with SNS junctions~\cite{Ridderbos2020HardNanowires}. On the left side of the QPC there is a PtSiGe superconducting lead and on the right side a normal lead consisting of a 2DHG accumulated in the Ge QW. With the accumulation gate (AG) set at large negative voltages to populate the QW we apply a more positive voltage to the constriction gates (CG), creating a tunable barrier between the superconducting and the normal region. In Fig.~\ref{fig:SN}b we progressively decrease the barrier height (decreasing $V_\mathrm{CG}$) going from the tunneling regime, where conductance is strongly suppressed, to a more open regime where conductance approaches the single conductance quantum $G_0$. Line-cuts of the conductance color map are presented in the bottom panel of Fig.~\ref{fig:SN}c. In the tunneling regime, we observe a hard induced superconducting gap, characterised by a two orders of magnitude suppression of the in-gap conductance to the normal-state conductance, and the arising of coherence peaks at $e|V_\mathrm{SD}|\approx \Delta^*=\SI{70}{\micro eV}$. Fig.~\ref{fig:SN}b also shows that the induced superconducting gap varies with the constriction gate voltage. A possible explanation is that, upon increasing the density in the semiconductor nearby the junction, the coupling to the parent superconductor might vary, as also observed in other hybrid nanostructures~\cite{DeMoor2018ElectricNanowires}.

The evolution of the gap as a function of in-plane magnetic field ($B_{\parallel}$) shown in Fig.~\ref{fig:SN}c confirms that the gap remains hard for finite magnetic fields up to \SI{0.25}{T}, ultimately vanishing at $B_{\parallel}\approx 0.37$~T. The magnetic field evolution of the gap in all three directions matches the behaviour observed in the SNS-QPC (Supplementary Fig.~S7).

Finally, Fig.~\ref{fig:SN}d reports the conductance traces in tunneling regime for all the six measured devices (an overview of the geometries of these devices and the respective measurements are available in the Supplementary Fig.~S8, the conductance maps for all these devices are shown in Supplementary Fig.~S9). For all devices we observe suppression of conductance equal or larger than two orders of magnitude. At a quantitative level, the conductance traces of Fig~\ref{fig:SN}d are well fitted by the BTK theory~\cite{Blonder1982TransitionConversion} (Supplementary Fig.~S9) consistent with a hard induced superconducting gap free of subgap states~\cite{Heedt2021Shadow-wallDevices, Chang2015HardNanowires}. This finding is the signature of a robust process that yields a reproducible high-quality superconductor-semiconductor interface, overcoming a long-standing challenge for hybrid superconductor-semiconductor quantum devices in Ge.

\subsection{Superconducting quantum interference devices}

\begin{figure}[!ht]
	\includegraphics[width=85mm]{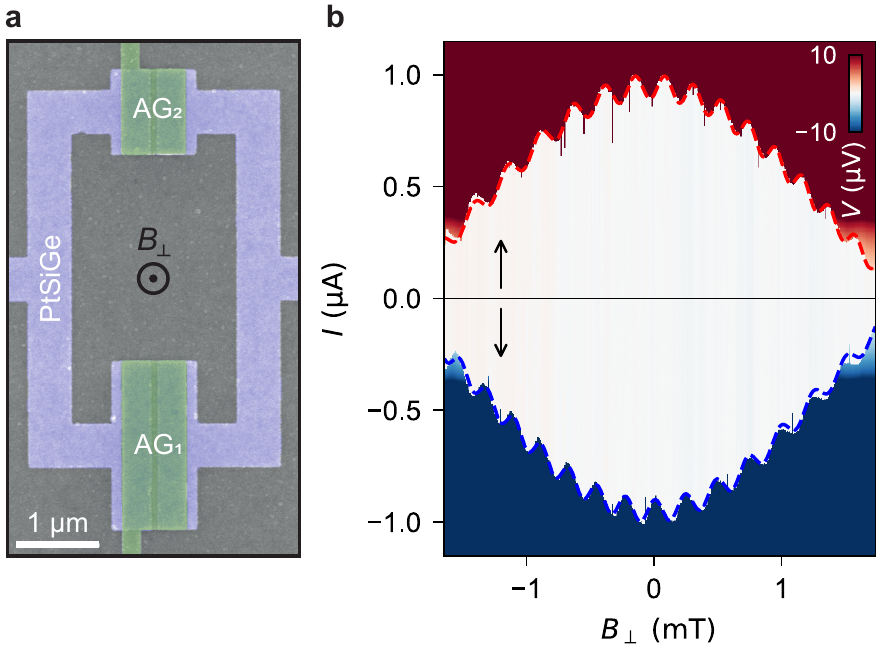}
	\caption{\textbf{Phase control of a Josephson junction in a SQUID. a}) False-color SEM image of the two JoFET SQUID device. The JoFETs have a channel length of \SI{70}{nm} and a channel width of \SI{1}{\micro m} and \SI{2}{\micro m} respectively and can be independently controlled by gates $\mathrm{AG}_1$ and $\mathrm{AG}_2$. The geometric loop area of the SQUID is of \SI{10}{\micro m^2}, calculated assuming a rectangle with sides positioned in the center of the PtSiGe loop cross-section.
	\textbf{b}) Color-plot of voltage drop ($V$) across the SQUID \emph{vs} current ($I$) and out-of-plane magnetic field ($B_{\perp}$). Arrows represent the direction of the current ($I$) sweep. With the gate voltages set at $V_\mathrm{AG1}= \SI{-3.5}{V}$ and $V_\mathrm{AG2}= \SI{-1.65}{V}$ the superconducting phase drops mainly over the second junction. Upon sweeping the out-of-plane magnetic field $B_{\perp}$ we observe oscillations of the switching current. Red and blue dashed lines are the fit of the evolution of the critical current with magnetic field. The magnetic field is applied in the out of plane direction as depicted in panel (a).}
\label{fig:squid}
\end{figure}

\begin{figure*}[!ht]
	\includegraphics[width=175mm]{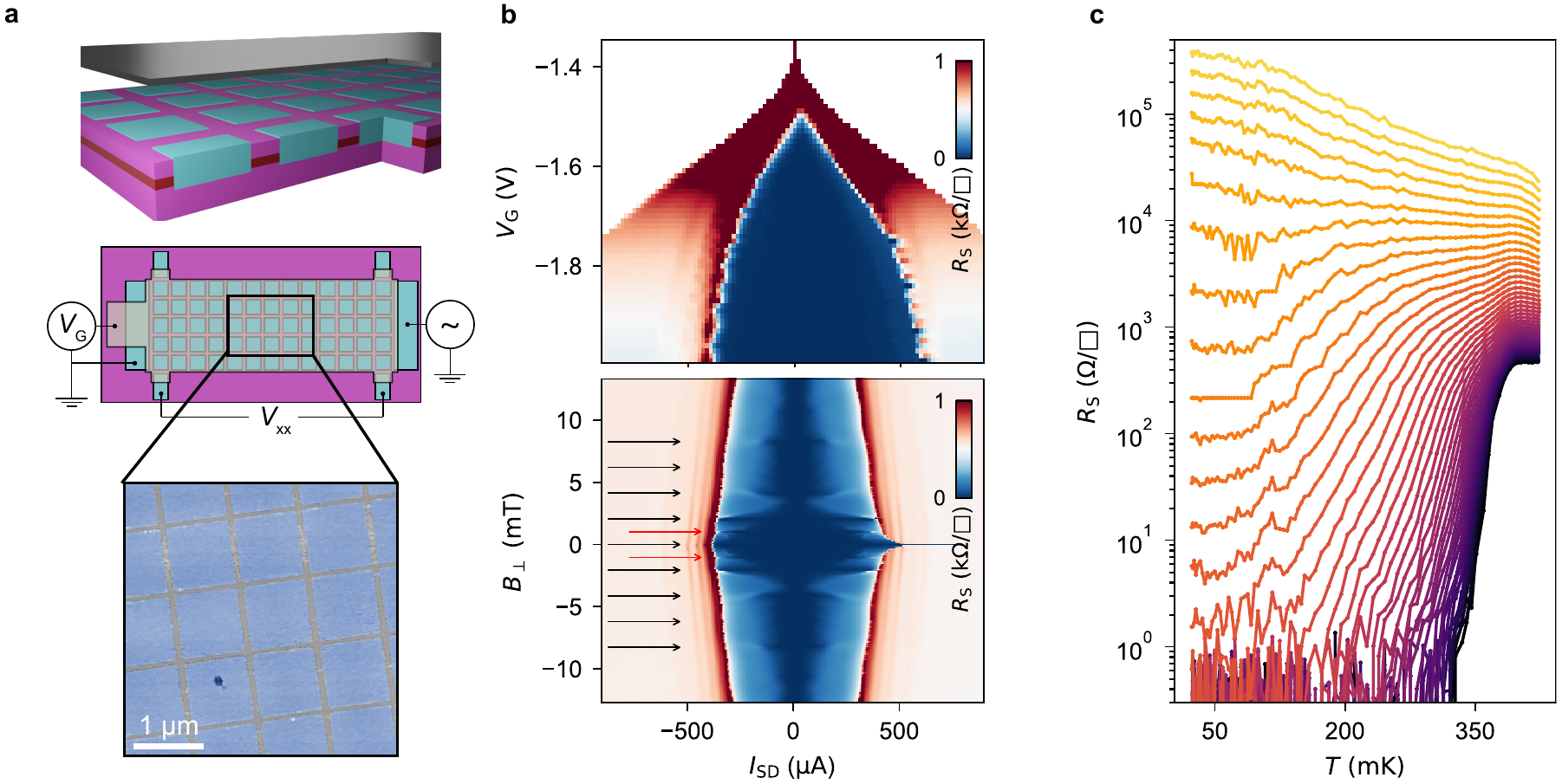}
	\caption{
	\textbf{A gated 2D superconductor-semiconductor array.} 
	\textbf{a}) 3D and top view schematics of an array of $51 \times 10$ PtSiGe islands. The inset shows an atomic force microscopy image of the PtSiGe islands of the array. The PtSiGe islands are $930\times930$~nm wide and the separation between neighbouring islands is of \SI{70}{nm}. 
	\textbf{b}) Top panel shows a color map of sheet resistance ($R_\mathrm{S}$) \emph{vs} accumulation gate voltage $V_\mathrm{G}$ and source-drain current $I_\mathrm{SD}$. Bottom panels shows a color map of sheet resistance \emph{vs} out of plane magnetic field $B$ and source-drain current $I_\mathrm{SD}$. The measurement is taken at gate voltage $V_\mathrm{G}=\SI{-1.99}{V}$, where we expect carriers in the quantum well to approach a saturation density value of about $6\times 10^{11}$~cm$^{-2}$ and have a mean free path ($\sim \SI{7}{\micro m}$) much longer than the separation between between neighbouring islands. Black arrows denote the magnetic field corresponding to one flux quantum $\Phi_0$ per unit cell of the array. Red arrows correspond to one-half flux per unit cell. 
	\textbf{c}) Sheet resistance as a function of temperature for gate voltages ranging from \SI{-2}{V} to \SI{-1.55}{V}. Yellow curves correspond to small negative gates, and purple curves to large negative gates.
	}
\label{fig:arrays}
\end{figure*}

We use the superconducting quantum interference device (SQUID) in Fig.~\ref{fig:squid}a to demonstrate phase control across a Josephson junction, an important ingredient for achieving topological states at low magnetic field~\cite{Pientka2017TopologicalJunction, Fornieri2019EvidenceJunctions, Ren2019TopologicalJunction,luethi_planar_2022}.
The device is composed of two Josephson field-effect transistors (JoFETs) with a width of \SI{2}{\micro\meter} and \SI{1}{\micro\meter} for JoFET$_1$ and JoFET$_2$ respectively, and equal length of \SI{70}{nm}. 
The critical current of the junctions $I_\mathrm{c1}$ and $I_\mathrm{c2}$ can be tuned independently by applying the accumulation gate voltages $V_\mathrm{AG1}$ and $V_\mathrm{AG2}$ to the corresponding gates. We investigate the  oscillations of the SQUID switching current as a function of the out-of-plane-magnetic field penetrating the SQUID loop. Namely, we set $V_\mathrm{AG1}$ and $V_\mathrm{AG2}$, such that both arms support supercurrent and $I_\mathrm{c1}\gg I_\mathrm{c2}$. This condition provides that the first junction is used as a reference junction and that the phase drop on it is flux independent, while the phase drop over the second junction is therefore modulated by the external flux through the loop. This allows the measurement of the current-phase-relation (CPR) of the second junction. This is demonstrated in Fig.~\ref{fig:squid}b where the shown SQUID oscillations are well fitted by the relation:
$I_\mathrm{c,SQUID}=I_\mathrm{c1}(B_{\perp} A_1)+I_\mathrm{c2}(B_{\perp} A_2) \sin(2 \pi B_{\perp} A_\mathrm{SQUID} - L I_\mathrm{c1}(B_{\perp} A_1)/\Phi_0)$ where $I_\mathrm{c1,2}(B A_{1,2})$ are the Fraunhofer dependencies of the critical current obtained from fitting the Fraunhofer pattern of each junction (Supplementary Fig.~S10), $A_{1,2}$ are the junction areas, $B_{\perp}$ is the out-of-plane magnetic field and $\Phi_0$ the flux quantum. From the fit of the data in Fig.~\ref{fig:squid}b (red dashed-line) we extract the effective SQUID loop area $A_\mathrm{SQUID}=\SI{8.9}{\micro m^2}$ (comparable to the \SI{10}{\micro m^2} SQUID geometric area) and the self-inductance $L=\SI{1.65}{pH}$. In order to confirm for the self-inductance effects, we also fit SQUID oscillations for the opposite direction of the current bias (blue dashed-line) and we get similar values for the effective loop area and self-inductance.

\subsection{Scalable junctions}
As a first step towards monolithic superconductor-semiconductor quantum circuits in two dimensions, we fabricate and study transport in a macroscopic hybrid device comprising a large array of 510 PtSiGe islands (Fig.~\ref{fig:arrays}a) and a global top gate. Each pair of neighbouring islands forms a Josephson junction whose transparency can be tuned by the global accumulation gate. The top panel of Fig.~\ref{fig:arrays}b shows a current bias measurement of the junctions array resistance. As the accumulation gate becomes more negative, all the junctions are proximitized and a supercurrent flows through the device. Remarkably, as the source-drain current approaches the junctions critical current the whole array simultaneously switches from superconducting to resistive regime, as shown from the sharp resistance step (Fig.~\ref{fig:arrays}b top). 

With this device we also study the evolution of the switching current in a small perpendicular magnetic field. In the bottom panel of Fig.~\ref{fig:arrays}b we observe Fraunhofer-like interference, along with the fingerprint of flux commensurability effects associated with the periodicity of the array. At integer numbers of flux quantum per unit area of the periodic array $f=B_{\perp}/B_0$, where $B_0=\Phi_0/A$ with $A$ the junction area and $\Phi_0$ the flux quanta, we observe switching current peaks at $\pm 1f$, $2f$, $3f$, $4f$ and $5f$, denoted by a black arrow in the plot. We also notice this effect at fractional values of $f$, most notably at $f/2$ (red arrow). Flux commensurability effects, due to the pinning and interference of vortices in Josephson junctions arrays, have been previously reported~\cite{Poccia2015CriticalTransition, Bttcher2018SuperconductingArray}. 

The observation of simultaneous switching of super-current and of the Fraunhofer pattern with flux commensurability effects, suggests that all islands effective areas are similar and that the supercurrent through the various junctions is comparable, meaning that all junctions respond synchronously to the applied gate voltage. This is further supported by the observation of sharp switching of super-current and the Fraunhofer pattern of a 1D array of superconducting islands presented in Supplementary Fig.~S11.

Finally we present in Fig.~\ref{fig:arrays}b the sheet resistance as a function of temperature for different gate voltages. As the gate voltage becomes more negative, the coupling between neighbouring superconducting islands increases and the system transitions from an insulating to a superconducting regime. At low gate voltage the resistance increases with decreasing temperature (yellow curves) indicating the insulating state, while at high gates the resistance drops to zero (purple curves) owing to the global superconducting state. At intermediate gate voltages ($\SI{-1.95}{V} \le V_\mathrm{G} \le \SI{-1.93}{V}$, orange curves) there is a transition where the resistance shows a weak temperature dependence. It will be interesting to study this regime in detail, in light of the recent claims of an anomalous metallic state between the superconducting and the insulating phases~\cite{Bttcher2018SuperconductingArray}.

\subsection{Conclusions}

In conclusion, we have developed superconducting germanosilicides for contacting Ge quantum wells, which has resulted in excellent superconducting properties imparted to the high-mobility 2DHG. We induced a hard superconducting gap in Ge, a large advancement compared to previous work on Ge hybrid superconductor-semiconductor devices~\cite{Hendrickx2019BallisticGermanium,Vigneau2019GermaniumInterferometers,Aggarwal2021EnhancementGas,Ridderbos2019HardNanowires}. We were able to observe a hard gap with 100\% yield across all the six measured devices, pointing to a robust and reproducible fabrication process. Next to this central result, we further demonstrate phase control across a Josephson junction and take advantage of the planar geometry to scale these devices in 2D arrays.

While we focused on the poly-crystalline superconducting PtSiGe compound, we anticipate two strategies to further increase the size of the induced superconducting gap, which sets a relevant energy scale for hybrid devices. Firstly, following the approach in ref.~\cite{Aggarwal2021EnhancementGas} a superconducting layer with a larger gap, such as Al or Nb, may be deposited on top of the superconducting PtSiGe.
Secondly, other ternary superconducting germanosilicides with a higher critical temperature may be explored, starting from the deposition and thermal anneal of other platinoid metals such as Rh and Ir~\cite{matthias_superconductivity_1963}.

Based on our findings, we foresee the following use cases for superconductor-semiconductor hybrids in high mobility planar Ge. With the challenge of a hard gap addressed, planar Ge appears a promising platform to host Majorana bound states in phased-biased Josephson junctions~\cite{hell_two-dimensional_2017,pientka_topological_2017,fornieri_evidence_2019}. Calculations with experimentally realistic material parameters~\cite{luethi_planar_2022} show that accessing the topological phase is feasible by careful design of Ge planar Josephson junctions geometries that relaxes magnetic field and spin-orbit constrains.

Crucially, the realization of a hard superconducting gap positions planar Ge as a unique material platform to pursue the coherent coupling of high fidelity spin qubits using crossed Andreev reflection to enable two-qubit gates over micrometer distances~\cite{choi_spin-dependent_2000,Leijnse2013CouplingSuperconductors}. Remote coupling of spin qubits in Ge may also be achieved by coupling spin qubits via superconducting quantum dots~\cite{choi_spin-dependent_2000,Leijnse2011QuantumSystems}, potentially offering a topological protection\cite{Kitaev2001UnpairedWires}. Coupling on an even longer distance may be obtained via superconducting resonators~\cite{Burkard2020SuperconductorsemiconductorElectrodynamics}. In such a scenario, a capacitive interaction may suffice, but connecting the resonator to a superconducting ohmic, such as PtSiGe, could result in a larger lever arm and therefore boost the coupling, while a direct tunnel coupling would give further directions to explore. The ability to couple qubits over different length scales is highly relevant and a critical component in network-based quantum computing~\cite{Vandersypen2017InterfacingCoherent}.

Furthermore, the demonstration of a hard gap in Ge motivates the investigation of alternative spin qubits systems, such as Andreev spin qubits (ASQ)~\cite{chtchelkatchev_andreev_2003,padurariu_theoretical_2010}, that may be coupled with gatemons~\cite{pita-vidal_direct_2022} or superconductors~\cite{spethmann_coupled_2022}. Similar to semiconductor spin qubits, the use of isotopically purified Ge~\cite{Itoh2014IsotopeApplications} may overcome the strong decoherence from the nuclear environment currently limiting progress with ASQs in III-V materials~\cite{hays_coherent_2021,pita-vidal_direct_2022}.

All together, these findings represent a major step in the Ge quantum information route, aiming to co-integrate spin, superconducting, and topological systems for scalable and high-fidelity quantum information processing on a silicon wafer.

\bibliography{references.bib}

\section{Methods}
\begin{footnotesize}
\textbf{Ge/SiGe heterostructure growth.} The Ge/SiGe heterostructure of this study is grown on a 100-mm n-type Si(001) substrate using an Epsilon 2000 (ASMI) reduced pressure chemical vapor deposition reactor. The layer sequence comprises a $\mathrm{Si_{0.2}Ge_{0.8}}$ virtual substrate obtained by reverse grading, a \SI{16}{nm} thick Ge quantum well, a \SI{22}{nm}-thick $\mathrm{Si_{0.2}Ge_{0.8}}$ barrier, and a thin sacrificial Si cap~\cite{Sammak2019ShallowTechnology}. Detailed electrical characterisation of heterostructure field effect transistors from these heterostructures are presented in ref.~\cite{Sammak2019ShallowTechnology}.
\\

\textbf{Device fabrication.} The fabrication of the devices presented in this paper entails the following steps. Wet etching of the sacrificial Si-cap in buffer oxide etch for \SI{10}{s}. Deposition of the Pt contacts via e-gun evaporation of \SI{15}{nm} of Pt at pressure of \SI{3e-6}{mbar} at the rate of \SI{0.5}{\angstrom /s}. Rapid thermal anneal of Pt contacts at \SI{400}{^\degree C} for \SI{15}{minutes} in a halogen lamps heated chamber in argon atmosphere. Atomic layer deposition of \SI{10}{nm} of $\mathrm{Al_2O_3}$ at \SI{300}{^\degree C}. Deposition of the first gate layer via e-gun evaporation of \SI{3}{nm} of Ti and \SI{17}{nm} of Pd. For the devices with a second gate layer the last two steps are repeated, \SI{27}{nm} of Pd are deposited for the second gate layer to guarantee film continuity where overlapping with first gate layer.
\\

\textbf{Transport measurements.}
Electrical transport measurements of the SNS-QPC, SN-QPC, SQUID devices are carried out in a dry dilution refrigerators at a base temperature of \SI{15}{mK}, corresponding to an electron temperature of $\approx$\SI{25}{mK} measured with a metallic N–S tunnel junction thermometer. This refrigerator is equipped with a 3-axis vector magnet. Measurements of the junctions array are carried out in a wet dilution refrigerator with base temperature of \SI{50}{mK} and z-axis magnet. 

Measurements are performed using a standard 4-terminals low-frequency lock-in technique at the frequency of \SI{17}{Hz}. Voltage bias measurements are performed with an excitation voltage $V_\mathrm{AC}<\SI{4}{\micro V}$. By measuring in a four-terminal setup, additional data processing to subtract series resistances of various circuit components is avoided. For the measurements in Fig.~\ref{fig:SNS}b, c, d and Fig.~\ref{fig:arrays}b the (maximum) gate voltage is tuned to be just below the threshold for hysteresis, caused by trapped charges in the surface states at the semiconductor/dielectric. In these electrostatic conditions the valence band edge at the semiconductor/dielectric interface and the Fermi level align and the density in the buried channel is expected to approach a saturation density of about $6 \times 10^{11}$~cm$^{-2}$~\cite{Sammak2019ShallowTechnology}.
\\

\textbf{Simulations and fitting of MARs.}
The experimentally measured conductance $G_{exp}(V)$ of an SNS junction is assumed to be superposition of N single-mode contributions~\cite{Kjaergaard2017TransparentJunction}:
\begin{equation}
    G_{theory}(V)\sum^M_{i=1} N_i G^{(\tau _i, \Delta)}(V) 
\end{equation}

where $G^{(\tau _i, \Delta)}$ is the simulated conductance for the $N_i$ modes with transparency $\tau_i$. We allow for $M$ different
transparencies, but all $N_i$ modes have the same superconducting gap $\Delta$. The simulations of conductance were implemented in Python using a modified version of the code presented in ref.~\cite{Nowak2019SupercurrentJunction}.

The theoretically computed conductance $G_{theory}(V)$ is fitted to $G_{exp}(V)$ using a nonlinear least-squares procedure: $\chi = \int [G_{exp}(V)-G_{theory}(V)]^2 dV$ is minimised for the fitting parameters $\Delta$, $N_i$, $\tau_i$ with $i \in {1,...,M}$. The fitting is performed for increasing $M$, provided that all $N_i$ and $\tau_i$ are nonzero. We note that we assume a coherent 1D system. When the MAR contribution is significant, this assumption leads to an overestimation of the sharpness and amplitude of the peaks. Nonetheless, overall we find a good agreement between the data and the model.
\\
\end{footnotesize}
\section{Acknowledgments}
We thank L. Kouwenhoven for fruitful discussions.

A. T. and G. S acknowledges support through a projectruimte associated
with the Netherlands Organization of Scientific Research (NWO). M. V. acknowledges support through an ERC Starting Grant.

ICN2 acknowledges funding from Generalitat de Catalunya 2017 SGR 327. ICN2 is supported by the Severo Ochoa program from Spanish MINECO (Grant No. SEV-2017-0706) and is funded by the CERCA Programme / Generalitat de Catalunya and ERDF funds from EU. Part of the present work has been performed in the framework of Universitat Autònoma de Barcelona Materials Science PhD program. Authors acknowledge the use of instrumentation as well as the technical advice provided by the National Facility ELECMI ICTS, node ”Laboratorio de MicroscopiasAvanzadas” at University of Zaragoza. M.B. acknowledges support from SUR Generalitat de Catalunya and the EU Social Fund; project ref. 2020 FI 00103. We acknowledge support from CSIC Research Platform on Quantum Technologies PTI-001. 

\section{Author contributions}
A.S. grew the Ge/SiGe heterostructures. A.T. fabricated the devices. C. N. B, M. B, S. M. , J. A. performed transmission electron microscopy characterisation. A. T. with C. B measured Josephson junction devices, SNS-QPCs and 2D arrays in wet dilution refrigerators, analysed the data and performed numerical simulation of the MARs processes. V. L. and A.T. measured SNS-QPC, SN-QPC devices and SQUIDs in dry dilution refrigerators and analyzed the data with the supervision of J. W. and input from F. B.. A.T. wrote the manuscript with input from all authors. G.S. conceived and supervised the project. 

\section{Additional information}

\section{Competing interests}

The authors declare no competing interests.

\section{Data availability}
All data included in this work are openly available at 4TU research data repository https://doi.org/10.4121/19940174.

\end{document}


\title{Supplementary Information: Hard superconducting gap in germanium}

\author{Alberto Tosato}
\affiliation{QuTech and Kavli Institute of Nanoscience, Delft University of Technology, PO Box 5046, 2600 GA Delft, The Netherlands}
\author{Vukan Levajac}
\affiliation{QuTech and Kavli Institute of Nanoscience, Delft University of Technology, PO Box 5046, 2600 GA Delft, The Netherlands}
\author{Ji-Yin Wang}
\affiliation{QuTech and Kavli Institute of Nanoscience, Delft University of Technology, PO Box 5046, 2600 GA Delft, The Netherlands}
\author{Casper J. Boor }
\affiliation{QuTech and Kavli Institute of Nanoscience, Delft University of Technology, PO Box 5046, 2600 GA Delft, The Netherlands}
\author{Francesco Borsoi}
\affiliation{QuTech and Kavli Institute of Nanoscience, Delft University of Technology, PO Box 5046, 2600 GA Delft, The Netherlands}
\author{Marc Botifoll}
\affiliation{Catalan Institute of Nanoscience and Nanotechnology (ICN2), CSIC and BIST, Campus UAB, Bellaterra, 08193 Barcelona, Catalonia, Spain}
\author{Carla N. Borja}
\affiliation{Catalan Institute of Nanoscience and Nanotechnology (ICN2), CSIC and BIST, Campus UAB, Bellaterra, 08193 Barcelona, Catalonia, Spain}
\author{Sara Martí-Sánchez}
\affiliation{Catalan Institute of Nanoscience and Nanotechnology (ICN2), CSIC and BIST, Campus UAB, Bellaterra, 08193 Barcelona, Catalonia, Spain}
\author{Jordi Arbiol}
\affiliation{Catalan Institute of Nanoscience and Nanotechnology (ICN2), CSIC and BIST, Campus UAB, Bellaterra, 08193 Barcelona, Catalonia, Spain}
\affiliation{ICREA, Passeig Lluís Companys 23, 08010 Barcelona, Catalonia, Spain}
\author{Amir Sammak}
\affiliation{QuTech and Netherlands Organisation for Applied Scientific Research (TNO), Stieltjesweg 1, 2628 CK Delft, The Netherlands}
\author{Menno Veldhorst }
\affiliation{QuTech and Kavli Institute of Nanoscience, Delft University of Technology, PO Box 5046, 2600 GA Delft, The Netherlands}
\author{Giordano Scappucci}
\email{g.scappucci@tudelft.nl}
\affiliation{QuTech and Kavli Institute of Nanoscience, Delft University of Technology, PO Box 5046, 2600 GA Delft, The Netherlands}

\date{\today}
\pacs{}

\maketitle
\tableofcontents
\newpage
\subsection{Two-dimensional hole gas properties}

\begin{figure*}[ht]
    \includegraphics[width=170mm]{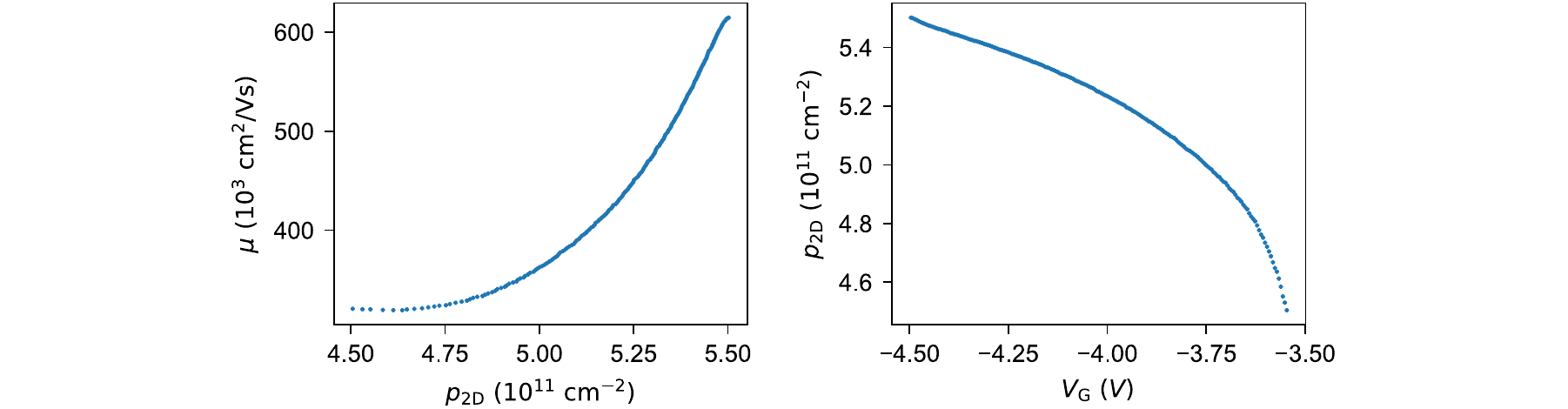}
    \caption{
    \textbf{2DHG transport properties.}
    Mobility $\mu$ vs 2D-carrier density $p_\mathrm{2D}$ (left panel) and 2D-carrier density  vs accumulation gate $V_\mathrm{G}$ for a Hall-bar shaped heterostructure field-effect transistor fabricated on the same \SI{22}{nm} deep Ge/SiGe heterostructure used for all devices in this work. The maximum mobility of \SI{615e3}{cm^2/Vs} is reached at the density of \SI{5.5e11}{cm^{-2}}, corresponding to an elastic transport scattering time $\tau = 31$~ps, calculated using $m^*=0.09$~\cite{Lodari2019LightWells} and a
    mean free path of \SI{7.4}{\micro m}.  The density vs gate curve deviates from the expected linear behaviour due to tunneling of charges from the quantum well to the the trap states at the oxide interface, partly screening the electric field in the quantum well. The density and mobility reach saturation when the states at the triangular well in the SiGe barrier at the oxide interface start to populate and thus screen the electric field in the QW.
    } 
    \label{fig:mob_dens}
\end{figure*}

\subsection{PtSiGe properties}

\begin{figure*}[ht]
    \includegraphics[width=160mm]{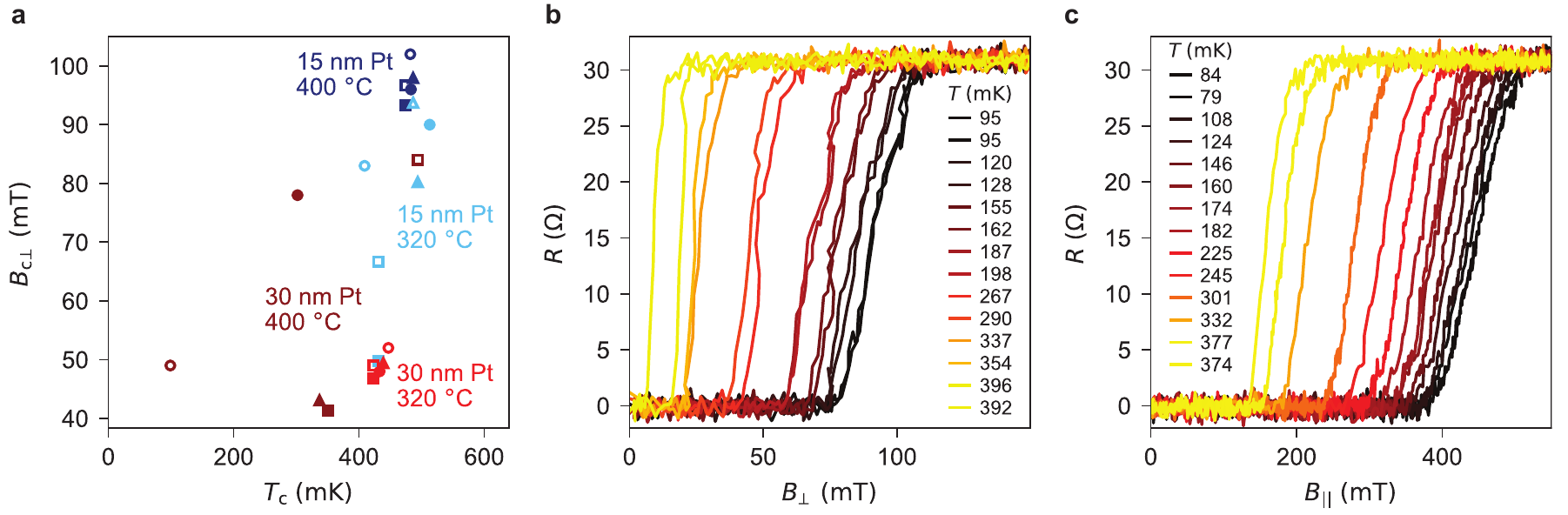}
    \caption{
    \textbf{PtSiGe film characterization.}
    \textbf{a}) Critical perpendicular magnetic field $B_\mathrm{c \perp}$ and critical temperature $T_\mathrm{c}$ of a PtSiGe film deposited and anneled in a \SI{22}{nm} deep Ge/SiGe QW for different process conditions. The colours of the markers indicate the thickness of the deposited platinum layer (that covers the whole surface of a $3\times 3$~mm
    Ge/SiGe heterostructure) and the anneal temperature. The filled (open) markers correspond to an anneal time of 15 (30) minutes. The marker’s shape signifies the used atomic layer deposition (ALD) of $\mathrm{Al_2O_3}$ process: no ALD (circles), ALD with 60 min pre-heating at 300 °C (squares), or ALD with 15 min pre-heating (triangles). In both ALD processes, 10 nm of $\mathrm{Al_2O_3}$ was deposited. 
    \textbf{b, c}) 
    Analysis of the critical temperature and fields of a \SI{3}{\micro m} wide PtSiGe strip (15 nm Pt has been annealed for 15 minutes at \SI{400}{\degree C}). Resistance $R$ versus perpendicular magnetic field ($B_\perp$) and parallel magnetic field ($B_\parallel$) for various temperatures $T$. 
    These measurements were performed in a 4-probe configuration with standard low frequency lock-in technique in a wet dilution refrigerator with electron temperature of \SI{100}{mK}.
    }
    \label{fig:PtSiGe_film_properties}
\end{figure*}
\newpage

\begin{figure*}[t]
    \includegraphics[width=160mm]{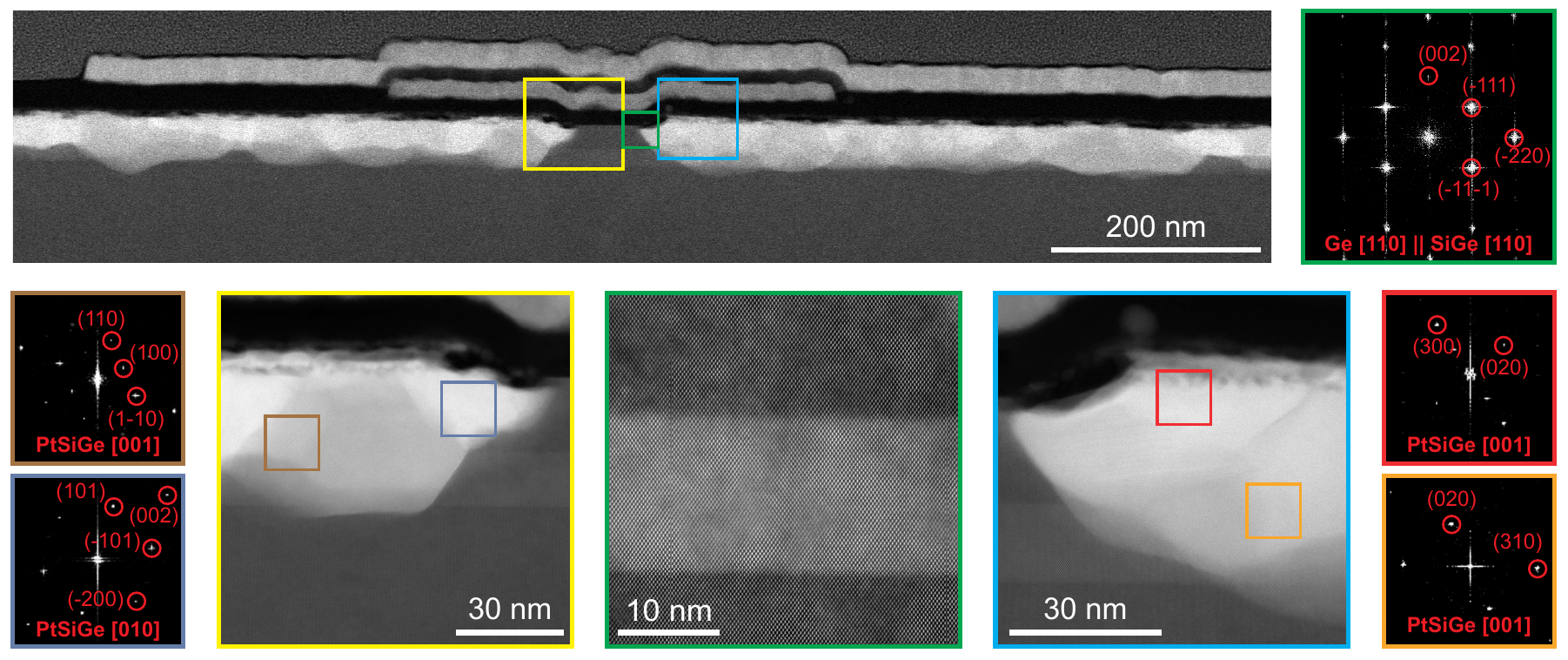}
    \caption{
    \textbf{Structural details of the PtSiGe poly-cristalline phase.} High-angle annular dark field scanning transmission electron microscopy (HAADF STEM) and crystallographic information of the SNS-QPC device. The yellow and blue insets show atomic-resolution images of both the left and right contacts highlighting the sharp interfaces between the QW and the PtSiGe film. The atomic-resolution micrograph in the center (green) displays the high quality of the Ge QW interfaces with diamond-structure (FD3-MS, space group number 227). The local contrast variations observed here are attributed to uneven thickness distribution of the lamella due to the focused ion beam (FIB) sample preparation. The fast Fourier transform (FFT) on the top right (green) indicates that the (002) planes in the QW grow epitaxially following the [001] axis. In addition, no dislocations were identified. The insets on the bottom left and right show the power spectra that identify the orthorhombic phase (PBNM, space group number 62) of the PtSiGe film.}
    \label{fig:}
\end{figure*}

\begin{figure*}[ht]
    \includegraphics[width=160mm]{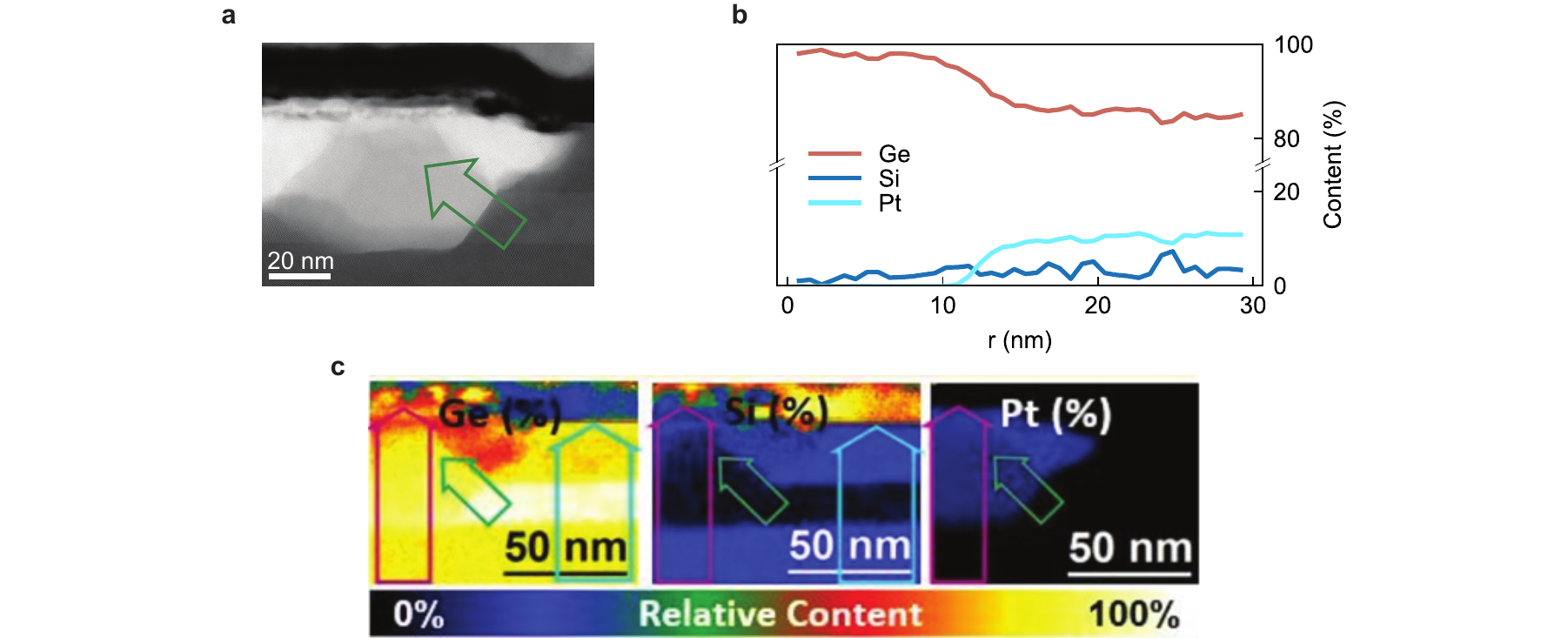}
    \caption{
    \textbf{PtSiGe stoichiometry.} Electron energy-loss spectroscopy (EELS) quantitative compositional map of the region indicated from the white arrow in the  HAADF STEM image (a) of the Ge/PtSiGe interface of the SNS-QPC. The threefold PtSiGe stoichiometry presented in panel (b) is Ge-rich, with relative composition in the range between $\mathrm{Pt_{0.1} Ge_{0.7} Si_{0.2}}$ and $\mathrm{Pt_{0.1} Ge_{0.85} Si_{0.05}}$ depending locally on the analysed grain. Panel (c) shows the quantitative EELS compositional maps for Ge Si and Pt. The averaged signal in the region along the green arrows is shown in panel (b).}
    \label{fig:EELS}
\end{figure*}

\newpage

\subsection{SNS-QPC measurements}

\begin{figure*}[ht]
    \includegraphics[width=145mm]{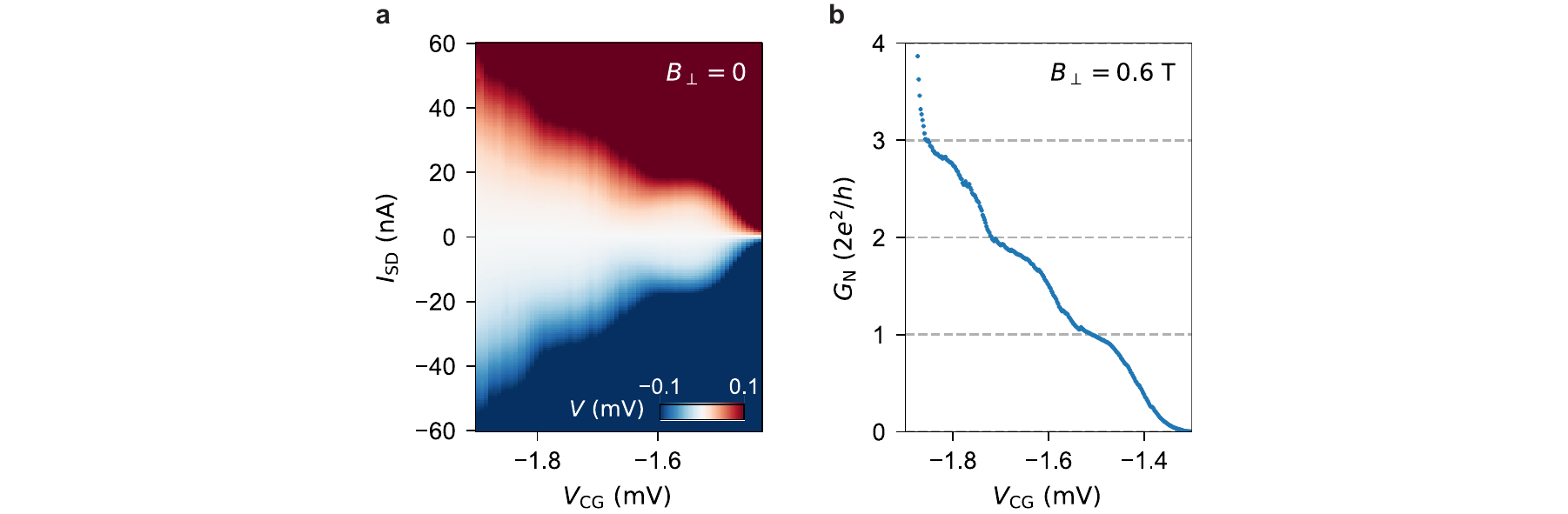}
    \caption{
    \textbf{Supercurrent discretization.}
    \textbf{a}) Voltage drop $V$ across an SNS-QPC device as a function of the source drain current $I_\mathrm{SD}$ and constriction gate voltage $V_\mathrm{CG}$. Discrete plateaus in the switching current can be observed, indicating a discrete number of modes in the QPC. 
    \textbf{b}) Normal-state differential conductance $G$ versus $V_\mathrm{CG}$ taken at out-of-plane magnetic field $B_{\perp}=\SI{0.6}{T}$, showing plateaus at quantized value of conductance. The plateaus in the two plots are slightly shifted with respect to each other due to the hysteretic behaviour of the device.
    }
    \label{fig:Ic_discretization}
\end{figure*}

\begin{figure*}[h]
    \includegraphics[width=145mm]{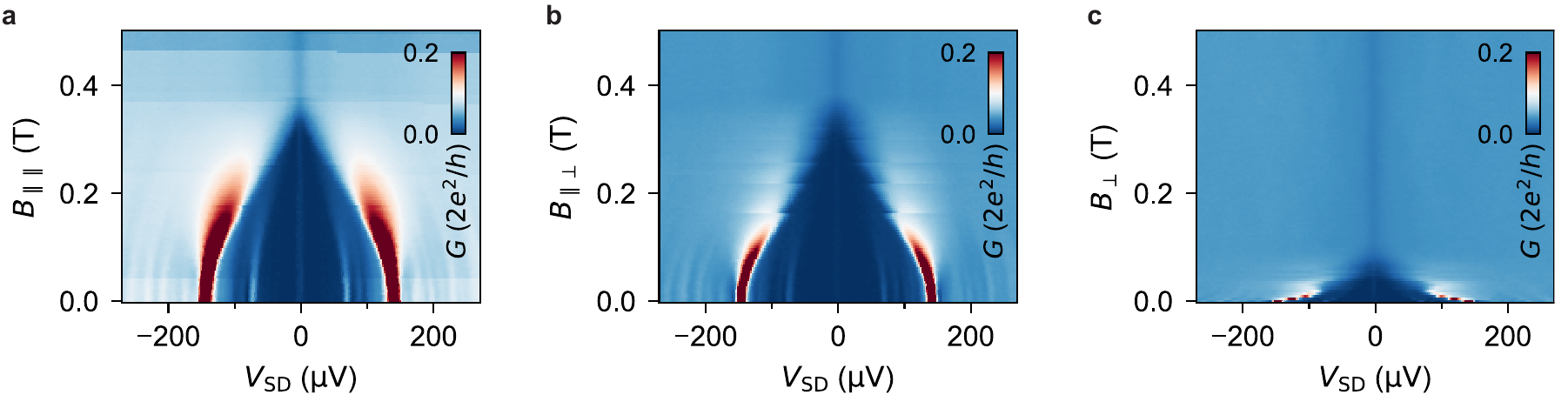}
    \caption{
    \textbf{SNS-QPC, evolution of the superconducting gap with magnetic field.} 
    Color map of conductance $G$ in units of $2e^2/h$ vs source-drain bias $V_\mathrm{SD}$ and magnetic field $B$ for the SNS-QPC. From left to right the magnetic field direction is: in-plane parallel to transport ($B_{\parallel \parallel}$), in-plane perpendicular to transport ($B_{\parallel \perp}$), out of plane ($B_{\perp}$). The device is tuned in the tunneling regime to show the evolution of the induced superconducting gap with the strength of the magnetic field.}
    \label{fig:SNS_field}
\end{figure*}

\subsection{SN-QPC measurements}

\begin{figure*}[h]
    \includegraphics[width=145mm]{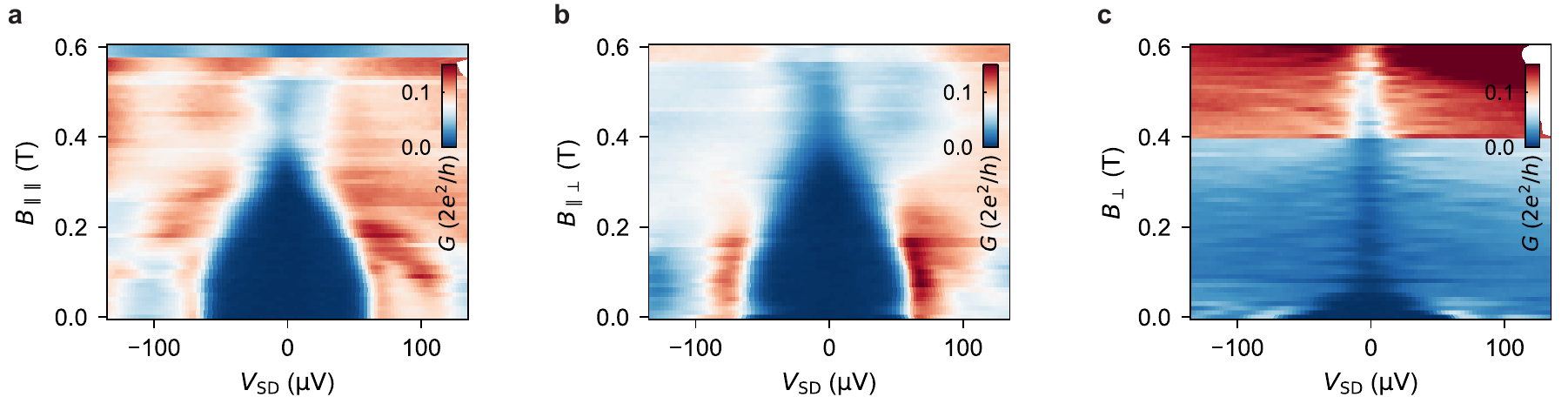}
    \caption{\textbf{SN-QPC, evolution of the induced superconducting gap with magnetic field.} Color map of conductance $G$ in units of $2e^2/h$ vs source-drain bias $V_\mathrm{SD}$ and magnetic field $B$ for the SN-QPC. From left to right the magnetic field direction is: in-plane parallel to transport ($B_{\parallel \parallel}$), in-plane perpendicular to transport ($B_{\parallel \perp}$), out of plane ($B_{\perp}$). The device is tuned in the tunneling regime to show the evolution of the induced superconducting gap with the strength of the magnetic field.}
    \label{fig:SN_field}
\end{figure*}

\newpage

\begin{figure*}[!htb]
    \includegraphics[width=160mm]{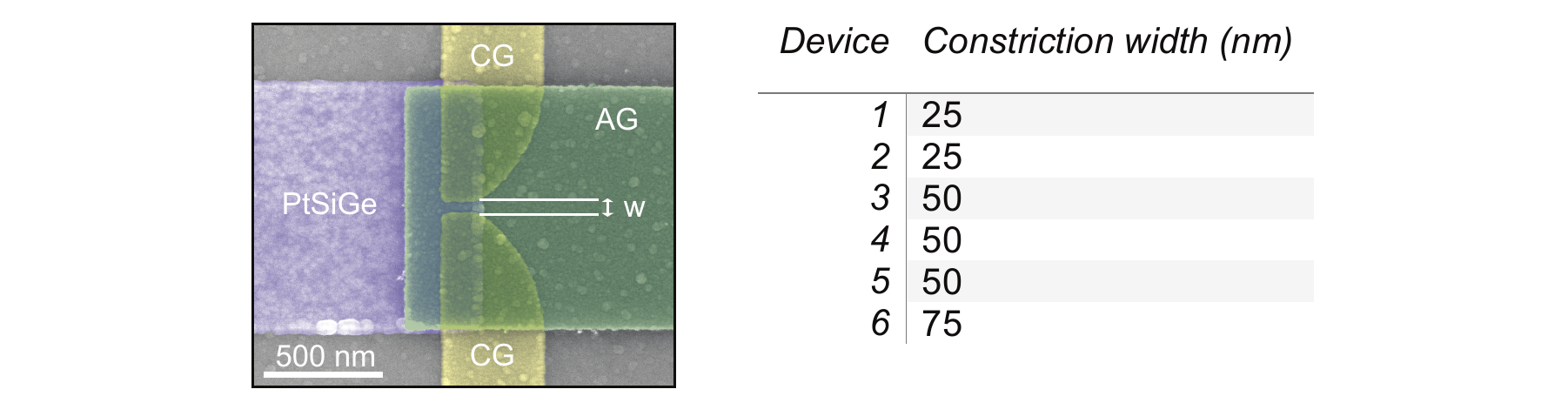}
    \caption{
    \textbf{SN-QPCs devices specifications.} False-color  SEM  image  of a superconductor-normal quantum point contact device (SN-QPC). The PtSiGe contact is violet, the constriction gates (CG) are yellow and the accumulation gate (AG) is green. The constriction width ($w$) between the two CGs is varied across the 6 measured devices and is reported in the table. The 6 devices were fabricated in the same fabrication run.}
    \label{fig:SN_devices_specs}
\end{figure*}

\begin{figure*}[!hbt]
    \includegraphics[width=145mm]{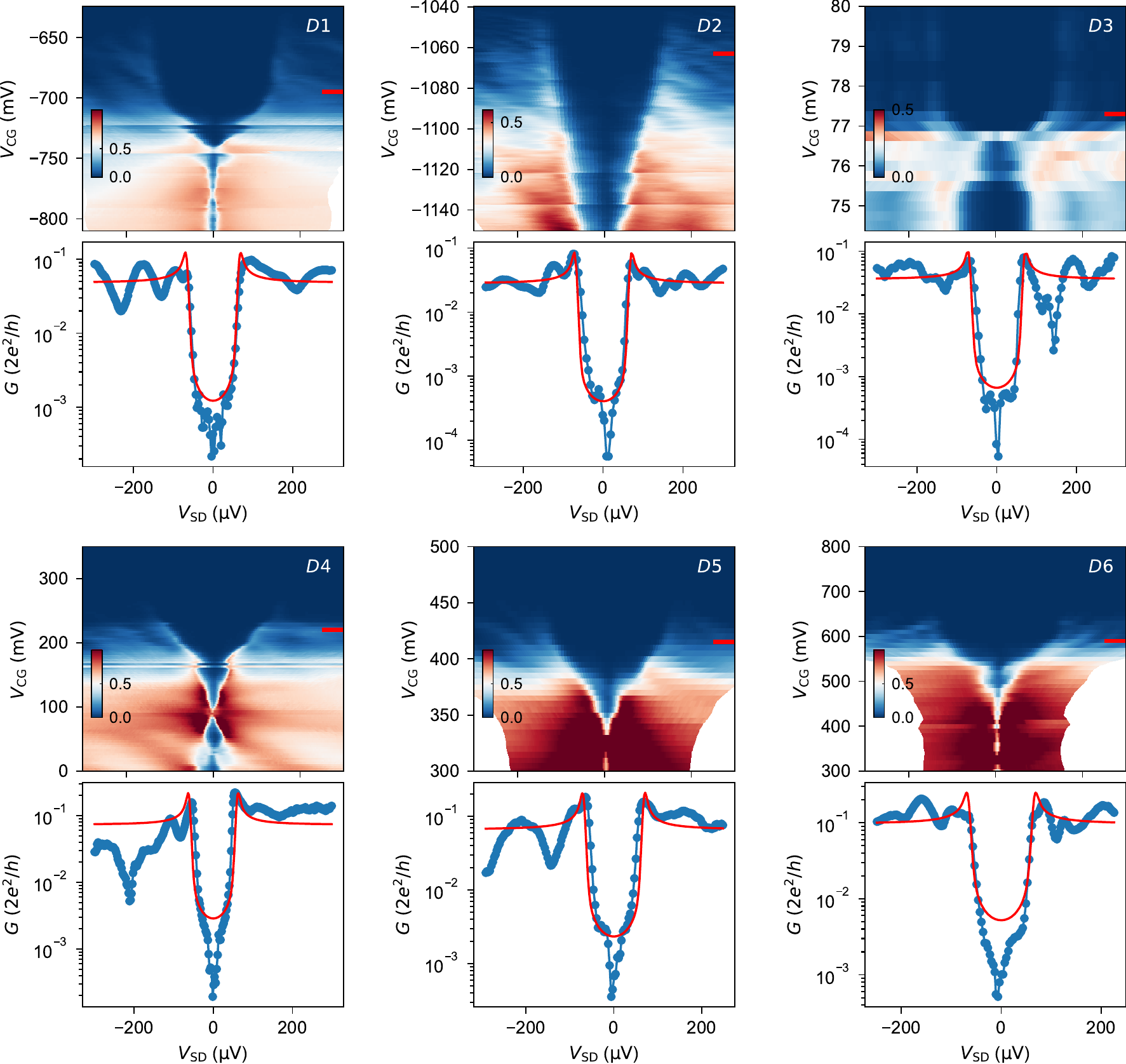}
    \caption{
    \textbf{Conductance maps of 6 SN-QPC devices.}
    Color map of $G$ in units of $2e^2/h$ vs. the source-drain voltage $V_\mathrm{SD}$ and constriction gate $V_\mathrm{CG}$, for the 6 SN-QPC devices presented in the main text, along with the conductance line-cuts presented in Fig.~3d main text. The red segment in the color maps indicates the $V_\mathrm{CG}$ of each linecut. Fits of the conductance linecuts to the BTK model \cite{Blonder1982TransitionConversion} (red lines) are consistent with a hard induced superconducting gap.  Variation on the $V_\mathrm{CG}$ operational window can be ascribed both to the different constriction gate size and to the accumulation gate voltage used for the specific measurement. The different evolution of $G$ as a function of $V_\mathrm{CG}$ can also be related to the different accumulation gate voltages. 
    }
    \label{fig:SN_G_maps}
\end{figure*}
\newpage

\subsection{SQUID measurements}

\begin{figure*}[!htb]
    \includegraphics[width=160mm]{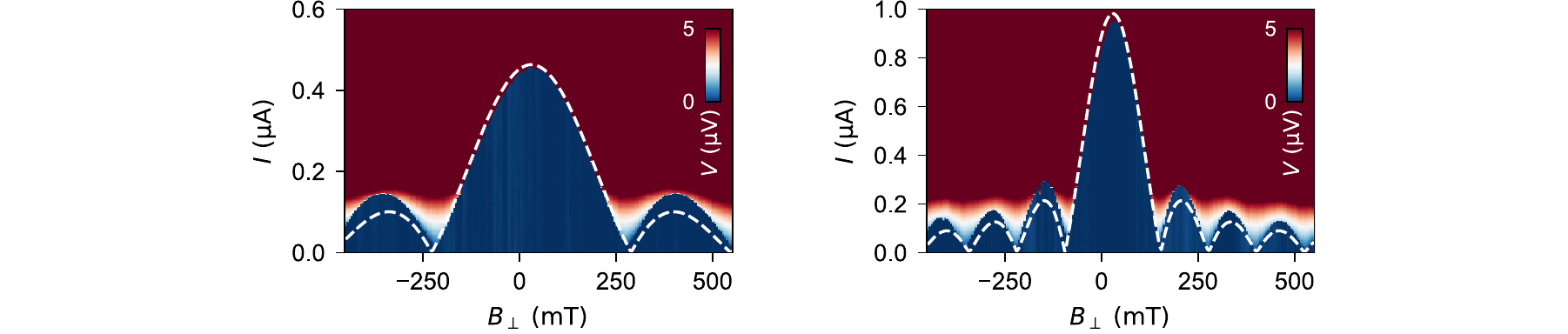}
    \caption{
    \textbf{JoFETs Fraunhofer pattern for the SQUID device.}
    Fraunhofer pattern of the small junction (JoFET$_2$, left panel) and large junction (JoFET$_1$, right panel) of the SQUID device. White dashed line represents the fitting of the switching current to the theoretical Fraunhofer formula.}
    \label{fig:squid}
\end{figure*}

We measured the Fraunhofer pattern for each junction of the SQUID device independently by measuring the dependence of its critical current on the out-of-plane magnetic field while the gate voltage of the measured junction is set to $\SI{-3.5}{V}$ and the other junction is pinched-off. By fitting the obtained dependencies $I_\mathrm{c1,2}(\Phi_{1,2})$ as $I_\mathrm{c1,2}(B A_{1,2})=I_\mathrm{c01,2} \sin(\pi B A_{1,2} /\Phi_0)/(\pi B A_{1,2}/\Phi_0)$, where $B$ is the out-of-plane magnetic field and $\Phi_0$ is superconducting flux quantum, we obtain from the fits the areas of the two junctions to be $A_1=\SI{1}{um^2}$ and $A_2=\SI{0.48}{um^2}$. Note that the ratio $A_1/A_2\sim 2$, as designed and shown in Fig. 4a, while the values for both areas are smaller than the geometrical areas in the design due to the flux focusing effects. 

\subsection{1D array}
\begin{figure*}[!htb]
    \includegraphics[width=160mm]{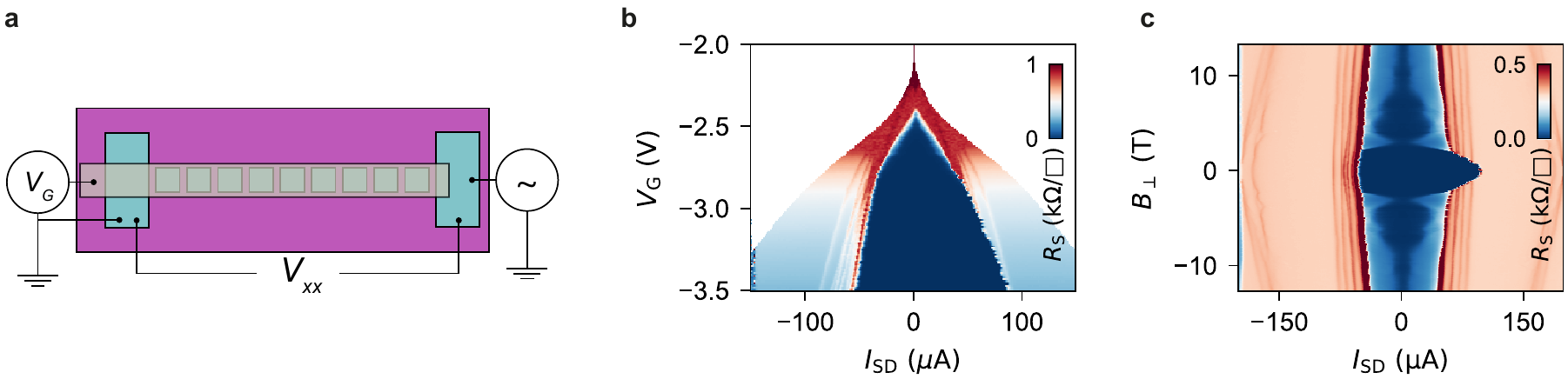}
    \caption{
    \textbf{1D PtSiGe superconducting array.}
    \textbf{a}) Top view schematics of an array of $51 \times 1$ PtSiGe islands on a Ge/SiGe heterostructure. The PtSiGe islands are $930\times930$~nm wide and the separation between the PtSiGe islands is of \SI{70}{nm}. \textbf{b}) Color map of the sheet resistance ($R_\mathrm{S}$) vs accumulation gate voltage $V_\mathrm{G}$ and source-drain current $I_\mathrm{SD}$. Increasing the negative voltage of the accumulation gate the array becomes superconducting ($R_\mathrm{S}$ goes to zero) when the source-drain current is below the switching current. \textbf{c}) Color map of the sheet resistance vs out-of-plane magnetic field $B_{\perp}$ and source-drain current $I_\mathrm{SD}$. The switching current shows the typical Fraunhofer pattern expected for a single Josephson junction. Compared to the 2D PtSiGe array this device does not present any signature of commensurability effects in the switching current, as expected for a linear array.
    }
    \label{fig:1D_array}
\end{figure*}
\newpage
\subsection{Key metrics}

\renewcommand{\arraystretch}{1.2}
\begin{table}[!ht]
    \centering
    \resizebox{17.5cm}{!}{\begin{tabular}{ |c|c||c|c|c|c|c|c|c|c| }

        \hline
        \multirow{2}{*}{\thead{Semiconductor}} & \multirow{2}{*}{\thead{Superconductor}} & $\mu$ & $\hbar/2\tau$ & $\Delta^* $  & $l_{SO}$ & $g^*$ & $T_1$ & $T_2^*$ & \thead{1Q gate fidelity} \\
        &&($\times$\SI{e3}{cm^2/Vs}) & (\SI{}{\micro eV}) & (\SI{}{\micro eV}) & (\SI{}{nm}) & & (\SI{}{ms}) & (ns) & (\%) \\
        \hline\hline
        Ge/SiGe, 2D & PtSiGe & 615 & 10  & 70 & 76 & 0.76-15 & 32 & 833 & 99.99 \\
        \hline
        InSb, nw & Al & 44 & 940 & 250  & 100 & 26-51 & na & 8 & na \\
        \hline
        \multirow{2}{*}{InAs, nw} & Al & \multirow{2}{*}{25} & 890 & 270  & \multirow{2}{*}{60} & \multirow{2}{*}{8} & \multirow{2}{*}{0.001} & \multirow{2}{*}{8} & \multirow{2}{*}{na}\\
        \cline{2-2} \cline{5-5}
         & Pb &&&1250&&&&& \\
        \hline
        InAs, 2D & Al & 60 & 370 & 190  & 45 & 10 &na&na&na\\
        \hline
        InSbAs, 2D & Al & 28 & 1200 & 220 & 60 & 55 & na & na & na\\ 
        \hline
    \end{tabular}}
    \caption{Comparison of key metrics for building quantum information processing devices based on topological or spin-qubit systems. We consider only systems where a hard hap is assessed via SN spectroscopy,  the most reliable measurement for verifying the absence of subgap states. From left to right, columns indicate: the semiconductor system and whether it is a planar heterostructures (2D) or a nanowire (nw); the superconductor material used to proximitize the semiconductor; maximum carrier mobility ($\mu$), typically Hall mobility in 2D systems and estimated field effect mobility in nanowires; the disorder quantified by the transport level broadening ($\hbar/2\tau$, where $\tau$ is the elastic scattering time)\cite{Ahn2021EstimatingNanowires}; maximum induced superconducting gap ($\Delta^*$); spin-orbit length ($l_{SO}$); g-factor ($g^*$), the range can be large when the g-factor is strongly anisotropic; longest relaxation time ($T_1$) measured in a spin qubit, longest dephasing time ($T_2^*$) measured in a spin-qubit; largest measured 1 qubit gate fidelity (1Q gate fidelity). The metrics reported in  this table are reported from the references below as following.
    Ge/SiGe, 2D: $\mu$ from this work; $\hbar/2\tau$ calculated using $m^*=0.09$~\cite{Lodari2019LightWells}; Ge/SiGe-PtSiGe $\Delta^*$ from this work; $l_{SO}=76$~nm (corresponding to a spin-orbit energy of \SI{2.2}{meV}) follows from the cubic Rashba coefficient $\alpha_3$ reported in ref.~\cite{DelVecchio2020VanishingGas} at a density of \SI{6.1E11}{cm^{-2}}, for which we assume an effective mass of 0.09~\cite{Sammak2019ShallowTechnology}; $g^*$~\cite{Mizokuchi2018BallisticGermanium}; $T_1$~\cite{Lawrie2020QuantumGermanium}; $T_2^*$~\cite{Hendrickx2020FastGermanium}; 1Q gate fidelity~\cite{Lawrie2021SimultaneousThreshold}.
    InSb, nw: $\mu$~\cite{Badawy2019HighNanowires}; $\hbar/2\tau$ calculated using $m^*=0.014$~\cite{kim2009accurate}; InSb-Al $\Delta^*$~\cite{OphetVeld2020In-planeNetworks}; $l_{SO}$~\cite{VanWeperen2015Spin-orbitNanowires}; $g^*$~\cite{Qu2016QuantizedContacts}; $T_2^*$~\cite{VanDenBerg2013FastNanowire}.; 1Q gate fidelity.
    InAs nw: $\mu$~\cite{Heedt2016BallisticContacts}; $\hbar/2\tau$ calculated using $m^*=0.0026$~\cite{kim2009accurate}; InAs-Pb $\Delta^*$~\cite{Kanne2021EpitaxialDevices}; InAs-Al $\Delta^*$~\cite{Deng2016MajoranaSystem}; $l_{SO}$~\cite{Liang2012StrongNanowires};  $g^*$~\cite{Heedt2016BallisticContacts}; $T_1$~\cite{Petersson2012CircuitQubit}; $T_2^*$~\cite{Nadj-Perge2010Spin-orbitNanowire}.
    InAs 2D: $\mu$~\cite{Aghaee2021InAs-AlProtocol}, $\hbar/2\tau$ calculated using $m^*=0.026$~\cite{kim2009accurate}, InAs-Al $\Delta^*$~\cite{Kjaergaard2016QuantizedHeterostructure}, $l_{SO}$~\cite{Shabani2016Two-dimensionalNetworks, Kjaergaard2016QuantizedHeterostructure},  $g^*$~\cite{Kjaergaard2016QuantizedHeterostructure}.
    InSbAs 2D: $\mathrm{\mu}$ \cite{Moehle2021InSbAsSuperconductivity}, $\hbar/2\tau$ calculated using $m^*=0.018$~\cite{Moehle2021InSbAsSuperconductivity}, $\Delta^*$~\cite{Moehle2021InSbAsSuperconductivity}, $l_{SO}$~\cite{Moehle2021InSbAsSuperconductivity}, $g^*$~\cite{Moehle2021InSbAsSuperconductivity}.
    }
    \label{tab:metrics}
\end{table}

In Table \ref{tab:metrics} we present a comparison of key metrics for material systems for hybrid superconductor-semiconductor applications. Given that in this paper the main focus is on applications that require the presence of a hard gap, we limit the table only to semiconductor-superconductor material systems with a hard gap assessed via SN spectroscopy, which is a reliable measurement for verifying the absence of subgap states. 

On the first half of the table we present the typical values for different platforms for (peak) mobility ($\mu$), disorder quantified by the transport level broadening ($\hbar/2\tau$, where $\tau$ is the elastic scattering time), size of induced superconducting gap ($\Delta^*$), spin orbit length ($l_{SO}$) and g-factor ($g^*$), important metrics for accessing the topological phase. On the second half of the table we illustrate the metrics that are significant for control and operation of spin qubits: relaxation time ($T_1$), dephasing time ($T_2^*$) and 1 qubit gate fidelity (1Q gate fidelity). For a comprehensive review of performance metrics of spin qubits in gated semiconducting nanostructures see ref.~\cite{stano_review_2022}. While III/V materials benefit from a larger induced superconducting gap and g-factor, planar Ge proximitized by PtSiGe stands out for the exceptionally low disorder (quantified by the high $\mu$ and low $\hbar/2\tau$), which is necessary for the emergence of topological Majorana zero modes~\cite{Ahn2021EstimatingNanowires}. In line with the remarks made in the introductory section, planar Ge also shows excellent spin qubit metrics. This comparison positions Ge/SiGe-PtGeSi as a compelling platform for topological devices, where small disorder is necessary to preserve the topological gap ( $\delta_\tau > \frac{\hbar}{2\tau}$, where $\delta_\tau$ is the topological gap and  and for hybrid devices, were we envision the coupling of spins via cross Andreev reflection mechanisms, Andreev spin qubits, and the co-integration of spins, topological, and superconducting qubits.

\bibliography{references_new.bib}